\documentclass{entcs}
\usepackage{entcsmacro}
\usepackage{graphicx}
\usepackage{kpfonts}
\usepackage{stmaryrd}
\usepackage{mathtools}
\usepackage{esvect}
\usepackage{url}
\usepackage{dsfont}
\usepackage{bbm}
\usepackage{braket}
\usepackage[inline]{enumitem}

\usepackage{tikz}

\usetikzlibrary{fit}
\usetikzlibrary{arrows,positioning} 
\tikzset{
    >=stealth',
    punkt/.style={
           rectangle,
           rounded corners,
           draw=black, very thick,
           text width=6.5em,
           minimum height=2em,
           text centered},
    pil/.style={
           ->,
           thick,
           shorten <=2pt,
           shorten >=2pt,}
}

\newenvironment{notation}{\medskip\refstepcounter{thm}\noindent\textbf{Notation \thethm~}\nopagebreak \begin{itshape}}{\end{itshape}\medskip}
\newenvironment{runningExample}{\medskip\refstepcounter{thm}\noindent\textbf{Running example \thethm~}\nopagebreak}{\medskip}

\definecolor{mydarkgreen}{cmyk}{0.85, 0.31, 0.96, 0.2}
\definecolor{myred}{cmyk}{0.08, 0.86, 0.75, 0.01}

\newcommand\mdoubleplus{\mathbin{+\mkern-7mu+}}

\def\lastname{Troj\'ak et al.}
\begin{document}
\begin{frontmatter}
\title{Executable Biochemical Space \\for Specification and Analysis of Biochemical Systems} 
\author{Matej Troj\'ak, David \v{S}afr\'anek, Lubo\v{s} Brim}
\address{Systems Biology Laboratory, Masaryk University, Brno, Czech Republic}
\author{Jakub \v{S}alagovi\v{c}, Jan \v{C}erven\'{y}}
\address{Global Change Research Centre AS CR, v. v. i., Brno, Czech Republic}

\let\thefootnote\relax\footnotetext{\textbf{Contact:} \href{mailto:xtrojak@fi.muni.cz}{xtrojak@fi.muni.cz} or \href{mailto:safranek@fi.muni.cz}{safranek@fi.muni.cz}}
\let\thefootnote\relax\footnotetext{This work has been supported by the Czech Science Foundation grant 18-00178S and Czech National Infrastructure grant LM2015055.}

\begin{abstract} 
We present the second generation of a rule-based language called Biochemical Space Language (BCSL) that combines the advantages of different approaches and thus makes an effort to overcome several problems with existing solutions. The key aspect of the language is the level of abstraction it uses, which allows scalable and compact hierarchical specification of biochemical entities. This abstraction enables unique analysis techniques to reason about properties of models written in the language on the semantic and syntactic level.
\end{abstract}
\end{frontmatter}
\section{Introduction}\label{intro}

Modelling complex systems in systems biology has to be conducted at
several levels of abstraction that reflect well the known
information~\cite{Kitano}. At every level, the system has to be
described rigorously in a formal language that allows avoiding
misunderstood and ambiguous interpretations. The more complex the
system is, the harder it is to describe it rigorously while not losing
human-readability and compactness of the description at the same time.
A modern biochemical system specification
language that can be sufficiently employed in systems biology practice
has to be \emph{hierarchical} and \emph{executable}. Hierarchical
description allows expressing individual system components at different levels of
detail. Since not all biochemical structures are known in detail, the
language has to support the expression of partial knowledge. On the other
end, executability allows automatic assigning the description with
appropriate formal (mathematical or programming) structures that enable simulation and exhaustive analysis of desired properties or revealing bugs in the description. 

Traditional approaches used to describe biochemical systems are: (i) a \emph{chemistry approach} employing ``mechanical'' descriptions by chemical reactions or (ii) a \emph{mathematical approach} using ordinary differential equations or other mathematical formalisms.
The problem of both approaches is \emph{scalability} in the description of the model and in its execution: even when the formulation of a model does not run into scalability issues, the execution or simulation might still be infeasible~\cite{Romers107136}.
To that end, computer science offers a \emph{computational approach} based on abstract languages with a variety of rigorous executable semantics. 
Relations among these approaches have been discussed in~\cite{Cardelli} and~\cite{Henzinger}. 

A promising computational approach is provided by \emph{rule-based modelling}~\cite{kappa_formal,BNGL} and process-algebraic frameworks~\cite{Cardelli,BioPEPA,BioSPI}. Rule-based models make a natural extension of the mechanical reaction-based models used in chemistry. Instead of operating with objects, rule-based frameworks operate with \emph{types} that allow avoiding the combinatorial explosion that occurs when underlying objects are specified directly. The semantics of the model is given in terms of \emph{rules} defined on given types. An important advantage of rule-based approach is that mathematical models can be automatically generated from them. In particular, instead of relying on a single mathematical formalism, different mathematical models can thus be obtained for a given model (e.g., ODEs~\cite{KaDE}, PDEs~\cite{Smoldyn}, chemical master equation or continuous-time Markov chains~\cite{Pauleve2010,sneddon2011efficient}, reaction-diffusion systems~\cite{So2013}, etc.).

Although rule-based models make a great alternative to mathematical models, they are not yet sufficiently used in practice. 
The reason is that existing formalisms rely on cryptic (symbolic) syntax and they are limited to a specific subset of interactions or are too abstract: BNGL~\cite{BNGL} and Kappa~\cite{kappa_formal} target protein-protein binding; BioSPI~\cite{BioSPI} and SPiM~\cite{SPiM} use very elemental asymmetric binary synchronisation primitives;
BioPEPA~\cite{BioPEPA} adapts process-algebraic framework to chemical
reactions while relaxing the compactness of combinatorial
interactions; Chromar~\cite{honorato2017chromar} utilises functional
programming. These languages can be thus understood as low-level
formalisms that allow precise formal description and analysis of
biological processes. Several high-level frameworks have been developed based on
principles of these formalisms: rxncon~\cite{Romers107136} focuses on regulatory interactions and allows construction of rules from experimental evidence, LBS~\cite{Pedersen} and LBS-$\kappa$~\cite{LBSKappa} enrich rule-based framework with modularity, PySB~\cite{Lopez646} embeds Kappa and BNGL into Python, MetaKappa~\cite{Danos2009} extends Kappa language by hierarchical inheritance of agent sites, BioCHAM~\cite{BioCHAM}
explicitly separates rules from their mathematical semantics. None of
these frameworks provides a sufficiently universal solution for
description and annotation of heterogeneous biophysical processes
integrated at the cellular level. Apparently, different approaches
need to be combined accordingly to make a universal hierarchical
modelling and annotation base that supports executability. The work
presented in~\cite{Goksel} targets bringing annotation standards into rule-based frameworks.

On the other end, SBML multi~\cite{SBMLmulti} transfers rule-based
description into a universal XML format that fixes the hierarchical
structure of objects and modularity of rules. It moves the rule-based
paradigm towards a~standard technique of describing biological
systems. However, it does not directly solve the executability and
advanced analysis issues that make an important aspect of rule-based
frameworks.

Our long-term aim is the development of a general modelling framework~\cite{cs2bio2013,Trojak2016}. Together with general annotation format Biochemical Space~\cite{BCS}, it respects the need for maintaining existing ODE models but allows to align them with a mechanistic rule-based description that is understandable by biologists, compact in size, executable in terms of allowing basic analysis tasks ensuring consistency of the description, and provides links to existing bioinformatics annotation databases. Such a comprehensive solution allows supporting modellers effort in building mathematical models that have clear biochemical meaning and can be easily integrated. Moreover, mechanistic descriptions can be later used as computational models having all advantages of rule-based modelling. To that end, we have pioneered an idea of combining advantages of rule-based modelling with the simplicity of chemical reactions by introducing the first prototype of a high-level rule-based language called \emph{Biochemical Space Language} (BCSL), introduced in~\cite{Ded201627}. The language has been defined at the top of Kappa. BCSL aims at higher-level abstraction than Kappa that focuses on morphisms between protein binding sites. Therefore the Kappa-based formulation of BCSL has limited expressiveness and does not fit well the aims of our framework. Additionally, Kappa does not provide hierarchical description which is one of the key aspects of BCSL.

In this paper, BCSL is redefined and significantly improved with respect to the primary prototype presented
in~\cite{Ded201627}: 
\begin{enumerate*}
\item hierarchical and composable object types and rules are defined without the need to encode them in an existing rule-based framework thus avoiding any loss of information,
\item executable semantics of rules is defined directly at the level of the language thus making a base for unique analysis tasks specific for the considered level of abstraction,
\item software tool is available to maintain and analyse BCSL specifications -- BCSgen\footnote{\href{https://github.com/sybila/BCSgen}{https://github.com/sybila/BCSgen}}.
\end{enumerate*}
The new version of BCSL emphasises the following aspects:
\begin{enumerate*}
\item \emph{human-readability} (easy to read, write, and maintain),
\item \emph{executability} (formal executable semantics is defined allowing efficient static analysis and consistency checking),
\item \emph{universality} (principally different cellular processes can be sufficiently combined in a single specification),
\item \emph{scalability} (combinatorial explosion of the description is avoided),
\item \emph{hierarchy} (object types are described hierarchically allowing compositional assembly from simpler structures).
\end{enumerate*}
Moreover,  we provide several static analysis techniques which take the
advantage from the specific level of abstraction. They
are aimed primarily at consistency checking, model reduction and reachability analysis. Particularly, \emph{rule redundancy elimination} allows detecting unnecessary rules in the models, \emph{context-based reduction} and \emph{static non-reachability analysis} uniquely deal with non-reachability in terms of preventing expensive transition system enumeration in cases when it is not necessarily needed. These techniques are demonstrated on a model of \emph{fibroblast growth factor} (FGF) signalling pathway and show practical impact in the field of static analysis.

\section{Formal definition of Biochemical Space Language}
\label{formal_definition}

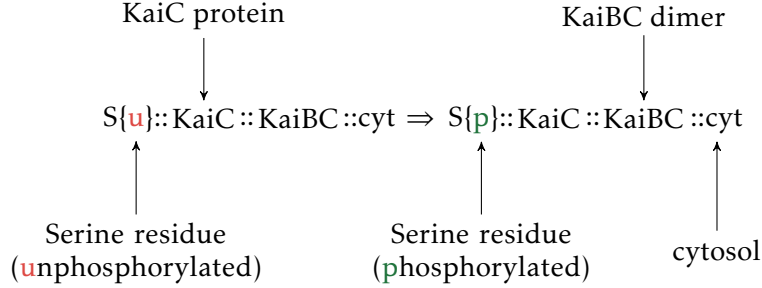
\begin{figure}[!h]
\begin{center}
\begin{tikzpicture}[scale=0.9,transform shape,  boxed/.style={rectangle, draw, rounded corners, minimum height=8em}]

    \node [font=\Large] (expSl) at (-9,27.98) {S\{\textcolor{myred}{u}\}::};
    \node [font=\Large] (expCl) at (-8,27.98) {KaiC};
    \node [font=\Large] (doublel) at (-7.36,27.98) {::};
    \node [font=\Large] (expBCl) at (-6.6,27.98) {KaiBC};
    \node [font=\Large] (expcytl) at (-5.55,27.95) {::cyt};
    \node [font=\Large, text width=4cm, align=center] (sl) at (-9,26)  {Serine residue (\textcolor{myred}{u}nphosphorylated)};
    \node [font=\Large] (kaicl) at (-8,29.5)  {KaiC protein};

    \node [font=\Large] (arrow) at (-4.8,27.98)  {$\Rightarrow$};

    \node [font=\Large] (expSr) at (-3.9,27.98) {S\{\textcolor{mydarkgreen}{p}\}::};
    \node [font=\Large] (expCr) at (-2.9,27.98) {KaiC};
    \node [font=\Large] (doubler) at (-2.25,27.98) {::};
    \node [font=\Large] (expBCr) at (-1.5,27.98) {KaiBC};
    \node [font=\Large] (expcytr) at (-0.42,27.95) {::cyt};
    \node [font=\Large, text width=4cm, align=center] (sr) at (-3.9,26)  {Serine residue (\textcolor{mydarkgreen}{p}hosphorylated)};
    \node [font=\Large] (kaibcr) at (-1.5,29.5) {KaiBC dimer};
    \node [font=\Large] (cytr) at (-0.42,26)  {cytosol};

	 \path[->]
	 (sl) edge [bend left=0] node [red!80, pos=0.5, sloped, above] {} (expSl)
	 (sr) edge [bend left=0] node [red!80, pos=0.5, sloped, above] {} (expSr)
     (kaicl) edge node [red!80, pos=0.5, sloped, below] {} (expCl)
     (kaibcr) edge node [blue, pos=0.4, sloped, above] {} (expBCr)
     (cytr) edge node [blue, pos=0.4, sloped, above] {} (expcytr);
\end{tikzpicture}
\caption{An example of a rule. The rule describes the change of serine (S) amino acid residue from an unphosphorylated to phosphorylated state. Additionally, such phosphorylation can happen only when the serine is part of a KaiC protein, which occurs inside a protein complex of KaiC and KaiB proteins. The entire process is allowed only inside of cytosol (cyt) compartment.}\label{rule_demo}
\end{center}
\end{figure}

In this section, we formally define Biochemical Space Language. At first, we define all the required objects (so called \emph{agents}) and interactions among them (so called \emph{rules}; for an example, see Figure~\ref{rule_demo}), then we define syntax of the language and semantics of the BCSL models.

\subsection{Formal preliminaries}

Before we proceed, we provide some basic definitions and notations in order to build the formal definition for the language.

\begin{definition}{(\emph{Multiset})}
\emph{Multiset} $\Omega$ is a pair $(\mathcal{A}, \mathtt{m})$ where $\mathcal{A}$ is a set and $ \mathtt{m} : \mathcal{A} \rightarrow \mathbb{N} $ is a function from $\mathcal{A}$ to the set of natural numbers. The set $\mathcal{A}$ is called the \emph{reference set} of elements. For each element $\mathit{a}$ in $\mathcal{A}$ the \emph{multiplicity} (that is, number of occurrences) of $\mathit{a}$ is the number $\mathtt{m}(\mathit{a})$.
\end{definition}

\begin{notation}
\begin{itemize}
  \item Let $\mathcal{S}$ be a set. By $\Omega^\mathcal{S}$ we denote the set of all possible finite multisets $(\mathcal{A}, \mathtt{m})$ such that $\mathcal{A} \subseteq \mathcal{S}$.
  \item Let $O = (o_1, \ldots, o_n)$ be a tuple.
  \begin{itemize}
    \item By $\Omega(O)$ we denote a multiset constructed from tuple $O$.
    \item By $\sigma(O)$ we denote a set of all possible permutations of length \emph{n} of the tuple $O$.
  \end{itemize}
  \item By $|Y|$ we denote 
  \begin{enumerate*}
    \item dimension of tuple $Y$ or
    \item cardinality of (multi)set $Y$.
  \end{enumerate*}
\end{itemize}
\end{notation}

\begin{definition}{(\emph{Labelled transition system})}\label{lts}
\emph{Labelled transition system} (LTS) $\mathcal{L}$ is a quadruple $(S, A, T, s_0)$ where $S$ is a set of states, $A$ is a set of labels, $T \subseteq S \times A \times S$ is a transition relation, and $s_0 \in S$ is an initial state.
\end{definition}

\begin{definition}{(\emph{Path in LTS})}
Let $\mathcal{L} = (S, A, T, s_0)$ be an LTS. We define \emph{path} as a sequence of states $\mathtt{s}_1 \mathtt{s}_2 \mathtt{s}_3 \ldots $ such that $\forall \mathtt{s}_i, \mathtt{s}_{i+1} : (\mathtt{s}_i, \mathtt{a}, \mathtt{s}_{i+1}) \in T$ for some $\mathtt{a} \in A$.
\end{definition}

\begin{definition}{(\emph{Tuples concatenation})}
Let $X = (x_1, \ldots, x_n), Y = (y_1, \ldots, y_m)$ be two tuples for some $n, m \in \mathbb{N}$. \emph{Concatenation} of two tuples, written $X \mdoubleplus Y$, is defined as:
$X \mdoubleplus Y = (x_1, \ldots, x_n, y_1, \ldots, y_m)$.
\end{definition}

\begin{definition}{(\emph{Sum of concatenations})}
Let $T = (T_1, T_2, \ldots, T_n)$ be sequence of tuples for some $n \in \mathbb{N}$. \emph{Concatenation} of sequence of tuples $\mdoubleplus_{i=1}^n T_i$ is defined as:
$\mdoubleplus_{i=1}^n T_i = T_1 \mdoubleplus T_2 \mdoubleplus \ldots \mdoubleplus T_n $
\end{definition}

\subsection{Objects definition}
\label{agents}

Let $\mathcal{N}_{A}, \mathcal{N}_{T}, \mathcal{N}_{\delta}, \mathcal{N}_{c}$ be mutually exclusive finite sets of atomic names, structure names, states, and compartments respectively. Moreover, $\varepsilon$ is a reserved symbol and does not belong to any of these sets.

For better readability, we provide examples of syntax for the most important objects with their definitions. The formal definition of syntax and the relation to the objects are given below (Sections~\ref{syntax} and \ref{semantic_function}).

\subsubsection{Signature}

\begin{definition}{(\emph{Signature})}
\emph{Atomic signature} is a function $\Sigma_{\mathtt{A}} : \mathcal{N}_{A} \rightarrow 2^{\mathcal{N}_{\delta}}$ that associates each atomic name to a set of state names. Similarly, 
\emph{structure signature} is a function $\Sigma_{\mathtt{T}} : \mathcal{N}_{T} \rightarrow 2^{\mathcal{N}_{A}}$ that associates each structure name to a set of atomic names.
\end{definition}

Signatures define a set of allowed states for an atomic name and an allowed set of atomic names for a~structure name.
For example, $\big\{~ S \rightarrow \{u, p\}$, $Q \rightarrow \{a, i\} ~\big\}$ is an atomic signature and $\big\{~ KaiC \rightarrow \{S, Q\}, KaiB \rightarrow \emptyset ~\big\}$ is a structure signature.

\subsubsection{Atomic agent}

\begin{definition}{(\emph{Atomic agent})}
An \emph{atomic agent} $\mathtt{A}$ is a pair $(\eta, \delta)$ where $\eta \in \mathcal{N}_{A}$ is a name and $\delta \in \mathcal{N}_{\delta}~\cup~\{ \varepsilon \}$ is a~state. The name and the state of the agent $\mathtt{A}$ is usually denoted by $\eta(\mathtt{A})$ and $\delta(\mathtt{A})$, respectively.
\end{definition}

Atomic agents are the simplest objects used for describing biological entities. Each atomic agent has its name and state. Allowed set of admissible states for the atomic agent (with additional empty $\varepsilon$ state) is given by signature $\Sigma_{\mathtt{A}}(\eta)$.

\begin{definition}{(\emph{Equality relation of atomic agents})}
Let $\mathtt{A},~ \mathtt{A}'$ be atomic agents. $\mathtt{A}$ is \emph{equal} to $\mathtt{A}'$, written $\mathtt{A} = \mathtt{A}'$, \emph{iff} $\eta(\mathtt{A}) = \eta(\mathtt{A}') \wedge \delta(\mathtt{A}) = \delta(\mathtt{A}')$.
\end{definition}

Intuitively, the defined equality on atomic agents is an equivalence relation.

\begin{notation}
We use the symbol $\mathds{A}$ to denote the universe of all possible atomic agents.
\end{notation}

Atomic agents are usually used to express small biological entities which can change their state, for example, amino acids, small inorganic molecules, etc.
Examples of atomic agents are $\mathtt{A}_1 = (S, \textcolor{mydarkgreen}{u})$, written as $S\{\textcolor{mydarkgreen}{u}\}$, and $\mathtt{A}_2 = (Q, \textcolor{mydarkgreen}{\varepsilon})$, written as $Q\{\textcolor{mydarkgreen}{\varepsilon}\}$.
Note the meaning of $\varepsilon$ is the state is unknown or not important to be considered in a given context.

\begin{definition}{(\emph{Compatibility of atomic agents})}
Let $\mathtt{A}_1$, $\mathtt{A}_2$ be atomic agents. The agent $\mathtt{A}_1$ is \emph{compatible with} agent $\mathtt{A}_2$, written $\mathtt{A}_1 \lhd \mathtt{A}_2$, if either $\mathtt{A}_1 = \mathtt{A}_2$ or $\eta(\mathtt{A}_1) = \eta(\mathtt{A}_2) \wedge \delta(\mathtt{A}_1) = \varepsilon $.
\end{definition}

Compatibility of atomic agents is a key property defined between agents. An agent is compatible with anothet agent if they have the same name and they are in the same state or the first agent is in the unknown state. It provides a formal way to compare which agent is more detailed, i.e. its state is more specified.

\begin{definition}{(\emph{Fully specified atomic agent})}
Let $\mathtt{A} \in \mathds{A}$ be an atomic agent. We say the agent $\mathtt{A}$ is \emph{fully specified}, written $\triangle \mathtt{A}$, \emph{iff} $\forall \mathtt{A}' \in \mathds{A} $ such that $\mathtt{A}' \neq \mathtt{A}: \neg (\mathtt{A}' \lhd \mathtt{A})$.
\end{definition}

\subsubsection{Structure agent}

\begin{definition}{(\emph{Structure agent})}
We define a \emph{structure agent} $\mathtt{T}$ as a pair $(\eta, \gamma)$ where $\eta \in \mathcal{N}_{T}$ is a name and $\gamma \subseteq \mathds{A}$ is a set of atomic agents called partial composition such that $\forall \mathtt{A}, \mathtt{A}' \in \gamma : \eta(\mathtt{A}) \neq \eta(\mathtt{A}')$. The name and the partial composition of the agent $\mathtt{T}$ is usually denoted by $\eta(\mathtt{T})$ and $\gamma(\mathtt{A})$, respectively.
\end{definition}

A structure agent represents a biochemical object that is composed of several known atomic agents while we know that a composition is abstract and not necessarily complete. To incorporate this kind of abstraction into our language, a structure agent is defined to be labelled with a unique name and a set of atomic agents. This set is restricted according to the given structure signature with the same name as the structure agent.

\begin{definition}{(\emph{Equality relation of structure agents})}
Let $\mathtt{T}, \mathtt{T}'$ be structure agents. $\mathtt{T}$ is \emph{equal} to $\mathtt{T}'$, written $\mathtt{T} = \mathtt{T}'$, \emph{iff} $\eta(\mathtt{T}) = \eta(\mathtt{T}') \wedge \gamma(\mathtt{T}) = \gamma(\mathtt{T'})$.
\end{definition}

Intuitively, the defined equality on structure agents is an equivalence relation.
The key construct of a structure agent is \emph{partial composition} defined as a set of atomic agents which are considered to be relevant parts of the structure agent. We allow this set to be empty with the meaning of a biological structure for which we do not know its composition.

\begin{notation}
We use symbol $\mathds{T}$ to denote the universe of all possible structure agents.
\end{notation}

A typical example of a structure agent is a protein where the atomic agents are amino acids that are of interest in the particular setting. Imagine that in our modelled system only three out of a few hundred amino acids are able to undergo some post-translational modifications, such as phosphorylation, metylation etc. It is suitable to model only these three amino acids instead of entire primary structure of the protein.
Examples of structure agent are $\mathtt{T}_1 = (K, \textcolor{myred}{\{}(S, \textcolor{mydarkgreen}{p}), (Q, \textcolor{mydarkgreen}{i})\textcolor{myred}{\}})$, written as $K\textcolor{myred}{(}S\{\textcolor{mydarkgreen}{p}\}, Q\{\textcolor{mydarkgreen}{i}\}\textcolor{myred}{)}$, and $\mathtt{T}_2 = (K, \textcolor{myred}{\{}(Q, \textcolor{mydarkgreen}{a})\textcolor{myred}{\}})$, written as $K\textcolor{myred}{(}Q\{\textcolor{mydarkgreen}{a}\}\textcolor{myred}{)}$.

We define difference on the level of partial compositions of structure agents, which is necessary for definition of semantics below.

\begin{definition}{(\emph{Difference of partial compositions})}
Let $\gamma, \gamma'$ be partial compositions. We define \emph{difference of partial compositions} 
$\gamma \ominus \gamma' = \{ \mathtt{A} ~|~ \mathtt{A} \in \gamma \wedge \mathtt{A} \not\in \gamma \cap \gamma'\}$ 
where
$\gamma \cap \gamma' = \{ \mathtt{A} ~|~ \mathtt{A} \in \gamma \wedge \exists \mathtt{A}' \in \gamma' : \eta(\mathtt{A}') = \eta(\mathtt{A}) \}$.
\end{definition}

\begin{definition}{(\emph{Compatibility of structure agents})}
Let $\mathtt{T}_1, \mathtt{T}_2$ be structure agents. The agent $\mathtt{T}_1$ is \emph{compatible with} agent $\mathtt{T}_2$, written $\mathtt{T}_1 \lhd \mathtt{T}_2$, \emph{iff} either $\mathtt{T}_1 = \mathtt{T}_2$ or $\eta(\mathtt{T}_1) = \eta(\mathtt{T}_2) \wedge \forall \mathtt{A}_1 \in \gamma(\mathtt{T}_1) ~\exists \mathtt{A}_2 \in \gamma(\mathtt{T}_2)$ : $\mathtt{A}_1 \lhd \mathtt{A}_2$.
\end{definition}

Structure agents are compatible if it is possible to create pairs from atomic agents of composition of the first agent with the second ones such that these atomic agents are all unique. For such pairs, the agents in each pair must be compatible. It provides a formal way to compare which agent is more specified, i.e. particular states of atomic agents in partial composition are given or not.

\begin{definition}{(\emph{Fully specified structure agent})}
Let $\mathtt{T} \in \mathds{T}$ be a complex agent. We say the agent $\mathtt{T}$ is \emph{fully specified}, written $\triangle \mathtt{T}$, \emph{iff} $\forall \mathtt{T}' \in \mathds{T} $ such that $\mathtt{T}' \neq \mathtt{T}: \neg (\mathtt{T}' \lhd \mathtt{T})$.
\end{definition}

\subsubsection{Complex agent}

A complex agent represents a non-trivial composite biochemical object that is inductively constructed from already known biological objects. In rule-based languages, this is usually defined by introducing bonds between individual biochemical objects. In BCSL we abstract from the detailed specification of bonds and we rather assume a complex as a coexistence of certain objects in a particular group. Moreover, a complex agent resides in a compartment which gives it a spatial position.

\begin{definition}{(\emph{Complex agent})}
We define a \emph{complex agent} $\mathtt{X}$ as a pair $(\mu, \mathtt{com})$ where $\mu \in (\mathds{A} \cup \mathds{T})^n$ is a sequence of agents, $\mathtt{com} \in \mathcal{N}_{c}$ is a compartment, and $n \in \mathbb{N}$. The sequence and the compartment of the agent $\mathtt{X}$ is usually denoted by $\mu(\mathtt{X})$ and $\mathtt{com}(\mathtt{X})$, respectively.
\end{definition}

The key element of a complex agent is \emph{sequence} inductively constructed from existing agents. In contrast to partial composition in structure agent, we allow replication at the level of sequence (an agent of a certain name can appear more than once in a sequence). The order in the sequence is necessary to uniquely identify agents which are equal. On the other hand, when comparing two sequences, we do it regardless the order.

\begin{definition}{(\emph{Equality relation of complex agents})}\label{rule_equiv}
Let $\mathtt{X}, \mathtt{X}'$ be complex agents. $\mathtt{X}$ is \emph{equal} to $\mathtt{X}'$, written $\mathtt{X} = \mathtt{X}'$, \emph{iff} 
$\mathtt{com}(\mathtt{X}) = \mathtt{com}(\mathtt{X'}) \wedge \Omega(\mu(\mathtt{X})) = \Omega(\mu(\mathtt{X'}))$.
\end{definition}

Intuitively, the defined equality on complex agents is an equivalence relation. Example of a complex agent is $\mathtt{X} = \big(((K, \textcolor{myred}{\{}(S, \textcolor{mydarkgreen}{p}), (Q, \textcolor{mydarkgreen}{i})\textcolor{myred}{\}}), (S, \textcolor{mydarkgreen}{p})), cell \big)$, written as $K\textcolor{myred}{(}S\{\textcolor{mydarkgreen}{p}\}, Q\{\textcolor{mydarkgreen}{i}\}\textcolor{myred}{)}.S\{\textcolor{mydarkgreen}{p}\}::cell$.

\begin{notation}
We use the symbol $\mathds{X}$ to denote the universe of all possible complex agents.
\end{notation}

The complex agents encapsulate other agents -- an atomic or a structure agent cannot exist on its own (the case when only one item is in its sequence can occur). This guarantees each atomic and structure agent has indirectly given spatial location -- the compartment.

\begin{definition}{(\emph{Compatibility of complex agents})}
Let $\mathtt{X}_1$, $\mathtt{X}_2$ be complex agents. The complex agent $\mathtt{X}_1$ is \emph{compatible with} complex agent $\mathtt{X}_2$, written $\mathtt{X}_1 \lhd \mathtt{X}_2$, \emph{iff} either $\mathtt{X}_1 = \mathtt{X}_2$ or $\mathtt{com}(\mathtt{X}_1) = \mathtt{com}(\mathtt{X}_2) \wedge \exists \mu' \in \sigma(\mu(\mathtt{X}_2))$ such that $ \forall i \in [1, n] : \mu_i(\mathtt{X}) \lhd \mu'_i $, where $n$ is length of sequence which is the same for both sequences.
\end{definition}

Complex agents are compatible if there exists a permutation of the sequence of the first agent such that individual agents on the same position in both sequences are compatible. It provides a formal way to compare which agent is more specified.

\begin{definition}{(\emph{Fully specified complex agent})}
Let $\mathtt{X} \in \mathds{X}$ be a complex agent. We say the agent $\mathtt{X}$ is \emph{fully specified}, written $\triangle \mathtt{X}$, \emph{iff} $\forall \mathtt{X}' \in \mathds{X} $ such that $\mathtt{X}' \neq \mathtt{X}: \neg (\mathtt{X}' \lhd \mathtt{X})$.
\end{definition}

It worth noting that the complexes have no binding topology. While it provides many advantages, specifically when it comes to combinatorial explosion, it also has several drawbacks. The most important one is that we are not able to express structural modifications on the level of complexes. These have to be encoded using states.

\subsubsection{Rule}

Let us have a simple example of a rule:

\begin{center}
$K\textcolor{myred}{(}S\{\textcolor{mydarkgreen}{u}\}\textcolor{myred}{)}.B\textcolor{myred}{(}\varnothing\textcolor{myred}{)}::cyt \Rightarrow K\textcolor{myred}{(}S\{\textcolor{mydarkgreen}{p}\}\textcolor{myred}{)}::cyt + B\textcolor{myred}{(}\varnothing\textcolor{myred}{)}::cyt$.
\end{center}

\noindent This rule dissociates a complex of $K$ and $B$ (both structure agents) to two separate agents while the structure agent $K$ is changing the state of its atomic agent $S$ from $\textcolor{mydarkgreen}{u}$ to $\textcolor{mydarkgreen}{p}$. In order to describe the rule formally, we need to capture the relation between so-called \emph{left-hand side} (the part \emph{before} $ \Rightarrow$ symbol) and \emph{right-hand side} (the part \emph{after} $ \Rightarrow$ symbol). It is achieved by indexing the individual positions in the rule and creating index maps between them.

\begin{definition}{(\emph{Rule})}
\label{rule_definition}
We define a \emph{rule} $\mathtt{R}$ as a quintuple $(\chi, \omega, \iota, \varphi, \psi)$ where:
\begin{itemize}
\item $\chi \in \mathds{X}^n$ is a sequence of complex agents,
\item $\omega \in (\mathds{A} \cup \mathds{T})^m$ is a sequence of atomic and structure agents,
\item $\iota \in \{ 0, \ldots, n \}$ is an index determining the end of the \emph{left-hand side} ($\mathit{LHS}$) of $\chi$,
\item $\varphi \in \mathbb{N}^m$ is an index map from $\omega$ to $\chi$,
\item $\psi \in ((\{-\} \cup \mathbb{N})^2)^n$ is an index map from $\mathit{LHS}$ to $\mathit{RHS}$
\end{itemize}

where $n, m \in \mathbb{N}$, $\mathit{LHS} = (\chi_1, \ldots, \chi_\iota)$ is the \emph{left-hand side}, and $\mathit{RHS} = (\chi_{\iota + 1}, \ldots, \chi_n)$ is the \emph{right-hand side}.
\end{definition}

The reason for this particular definition is that it is necessary to capture the relationship between the left-hand side and the right-hand side of the rule. This is done by enumerating all atomic and structure agents $\omega$ from sequence of complex agents $\chi$. The index map $\psi$ between the agents in $\omega$ determines pairs of agents from the left-hand side and the right-hand side which correspond to each other. It is possible that there are agents which do not have a pair (denoted by $-$) in the situation when the rule is modelling \emph{inflow} from (resp. \emph{outflow} to) the system. Another index map $\varphi$ serves for relating agents from $\omega$ back to the original sequence of complexes $\chi$. Finally, by index $\iota$ we determine the end of the left-hand side of the rule. Note the index is zero in the situation when there are no agents on the left-hand side.

\begin{notation}
We use symbol $\mathds{R}$ to denote the universe of all possible rules.
\end{notation}

Example of a rule is $\mathtt{R} = (\chi, \omega, \iota, \varphi, \psi)$ where:

\begin{minipage}{0.45\textwidth}

\begin{itemize}
\item $\chi = \begin{bmatrix}
\big(( (K, \textcolor{myred}{\{}(S, \textcolor{mydarkgreen}{u})\textcolor{myred}{\}}), (B, \textcolor{myred}{\emptyset}) ), cyt\big),\\

\big(((C, \textcolor{myred}{\emptyset}), (D, \textcolor{mydarkgreen}{i} ) ), cyt\big),\\

\big(((A, \textcolor{mydarkgreen}{\varepsilon} ) ), cyt\big),\\

\big(((K, \textcolor{myred}{\{}(S, \textcolor{mydarkgreen}{p} )\textcolor{myred}{\}}), (B, \textcolor{myred}{\emptyset}), (C, \textcolor{myred}{\emptyset}) ), cyt\big),\\

\big(((D, \textcolor{mydarkgreen}{a} ), (A, \textcolor{mydarkgreen}{\varepsilon} ) ), cyt\big),\\

\big(((H, \textcolor{mydarkgreen}{u}) ), cyt\big)  
\end{bmatrix}$
\end{itemize}

\end{minipage}%
\hfill
\begin{minipage}{0.45\textwidth}
\begin{itemize}
\item $\omega = \begin{bmatrix}
(K, \textcolor{myred}{\{}(S,  \textcolor{mydarkgreen}{u})\textcolor{myred}{\}}), (B, \textcolor{myred}{\emptyset}), (C, \textcolor{myred}{\emptyset}), \\
(D,  \textcolor{mydarkgreen}{i}), (A,  \textcolor{mydarkgreen}{\varepsilon}), (K, \textcolor{myred}{\{}(S,  \textcolor{mydarkgreen}{p})\textcolor{myred}{\}}), \\
(B, \textcolor{myred}{\emptyset}), (C, \textcolor{myred}{\emptyset}), (D,  \textcolor{mydarkgreen}{a}), (A,  \textcolor{mydarkgreen}{\varepsilon}), (H, \textcolor{mydarkgreen}{u})
\end{bmatrix}$

\item $\iota = 3$
\item $\varphi = (2,4,5,8,10,11)$
\item $\psi = [ (1,6) ; (2,7) ; (3,8) ; (4,9) ; (5,10) ; (-,11) ] $
\end{itemize}

\end{minipage}%

 written as:

\begin{center}
$K\textcolor{myred}{(}S\{\textcolor{mydarkgreen}{u}\}\textcolor{myred}{)}.B\textcolor{myred}{(}\varnothing\textcolor{myred}{)}::cyt + C\textcolor{myred}{(}\varnothing\textcolor{myred}{)}.D\{\textcolor{mydarkgreen}{i}\}::cyt + A\{\textcolor{mydarkgreen}{\varepsilon}\}::cyt \Rightarrow K\textcolor{myred}{(}S\{\textcolor{mydarkgreen}{p}\}\textcolor{myred}{)}.B\textcolor{myred}{(}\varnothing\textcolor{myred}{)}.C\textcolor{myred}{(}\varnothing\textcolor{myred}{)}::cyt + D\{\textcolor{mydarkgreen}{a}\}.A\{\textcolor{mydarkgreen}{\varepsilon}\}::cyt + H\{\textcolor{mydarkgreen}{u}\}::cyt$
\end{center}

Not every rule makes sense. For example, a rule where not a single agent is changed or a rule where the relation between the left-hand and the right-hand side would not be clear. In order to avoid such cases we need to specify when a rule is \emph{well-formed}, i.e. it makes sense semantically.

\begin{definition}{(\emph{Well-formed rule})}
\label{well_formed}
Let $\mathtt{R}$ be a rule and $i,j \in \mathbb{N}$. We say the rule $\mathtt{R} = (\chi, \omega, \iota, \varphi, \psi)$ is \emph{well-formed} if all the following conditions hold:

\begin{enumerate}
  \item \label{cond1} at least one of conditions holds:
  \begin{enumerate}
    \item \label{cond1a} $\exists (i,j) \in \psi: \omega_i \neq \omega_j $,
    \item \label{cond1b} $|\mathit{LHS}(\mathtt{R})| \neq |\mathit{RHS}(\mathtt{R})|$,
    \item \label{cond1c} $\exists i \in [1, \iota]: \mathtt{com}(\chi_i) \neq \mathtt{com}(\chi_{\iota + i})$;
  \end{enumerate}
  \item \label{cond2} $\forall (i,j) \in \psi: \eta(\omega_i) = \eta(\omega_j)$;
  \item \label{cond3} $\forall (-,i) \in \psi : ~\triangle \omega_i$.
\end{enumerate}

\end{definition}

A rule is well-formed if it holds conditions given in Definition~\ref{well_formed}. The conditions basically claim that an agent has to change during the rule application. This is ensured by condition~(\ref{cond1}), where there are three options: (\ref{cond1a}) at least one pair of agents from $\mathit{LHS}$ and $\mathit{RHS}$ of the rule is different; (\ref{cond1b}) the lengths of the $\mathit{LHS}$ and $\mathit{RHS}$ are different, i.e. either a new agent is created or complex is formed/dissociated; (\ref{cond1c}) a compartment is changed. Any combination of these sub-conditions is allowed.
The second condition~(\ref{cond2}) guarantees that the pairs of structure and atomic agents in $\omega$ of the rule have the same name.
Please note the conditions (\ref{cond1}) and (\ref{cond2}) do not apply to those agents in $\omega$ which do not have a pair on the other side of the rule.
Finally, the condition (\ref{cond3}) claims that if there is an agent which does not have defined a pair via index map $\psi$ (denoted by~$-$), it is required to be a fully specified agent (but only in case of agent creation, it is not necessary for agent degradation).

\subsection{Syntax}
\label{syntax}

In this section, we define the syntax for the language, i.e. how we usually write it in order to make the notation easily writeable and readable. It corresponds to the examples given while defining agents and rules above.

\begin{definition}{(\emph{Grammar})}
\label{grammar}
\begin{center}
\begin{tabular}{ l@{\hskip 2cm}l l }
Atomic expression & Structure expression & Complex expression \\
$\alpha ::= \eta\{s\} ~|~ \eta\{\varepsilon\}$ & $\tau ::= \eta(\gamma) ~|~ \eta(\varnothing)$ & $\Gamma ::= \beta_1~.~\ldots~.~\beta_k :: c$\\

$\eta ::= n \in \mathcal{N}_{A}$ & $\gamma ::= \alpha_1, \ldots, \alpha_k$ & $\beta_i ::= \alpha ~|~ \tau$ \\

$s ::= n \in \mathcal{N}_{\delta}$ & $\eta ::= n \in \mathcal{N}_{T}$ & $c ::= n \in \mathcal{N}_{c}$\\
 & & \\
\hline
 & & \\
Rule expression & $\varrho ::= \Gamma_1 + \ldots + \Gamma_n \Rightarrow \Gamma_{n+1} + \ldots + \Gamma_m $ & \\
 & & \\
\end{tabular}

\end{center}
where $m,n \in \mathbb{N}_0 \wedge m > n $ and $k \in \mathbb{N}$.
\end{definition}

\subsection{Translation function}
\label{semantic_function}

Once we defined BCSL agents and rules and syntax for the language, we need to connect them in order to give semantic meaning to a model written in the syntax. For this purpose, we define translation function $\mathtt{F}$ (Definition~\ref{translation_function}). It is defined recursively according to the expression given as an argument. 

\begin{definition}{(\emph{Translation function})}
\label{translation_function}
We define \emph{translation function} $\mathtt{F}$ according to the expression given in double square brackets $\llbracket ~\ldots~ \rrbracket$ as follows:

\begin{center}
\begin{tabular}{ r c l}
$\mathtt{F} \llbracket ~\eta\{\varepsilon\}~ \rrbracket$ & = & $(\eta, \varepsilon) \in \mathds{A}$\\
$\mathtt{F} \llbracket ~\eta\{s\}~ \rrbracket$ & = & $(\eta, s) \in \mathds{A}$\\
$\mathtt{F} \llbracket ~\eta(\varnothing)~ \rrbracket$ & = & $(\eta, \emptyset) \in \mathds{T}$\\
$\mathtt{F} \llbracket ~\eta(\mathtt{a}_1, \ldots, \mathtt{a}_k)~ \rrbracket$ & = &
$\big(\eta, \{~ \mathtt{F} \llbracket \mathtt{a}_1 \rrbracket, \ldots, \mathtt{F} \llbracket \mathtt{a}_k \rrbracket ~\}\big) \in \mathds{T}$\\
$\mathtt{F} \llbracket ~\alpha_1~.~\ldots~.~\alpha_k :: c~ \rrbracket$ & = &
$\big(~(\mathtt{F} \llbracket ~\alpha_1~ \rrbracket, \ldots, \mathtt{F} \llbracket ~\alpha_k~ \rrbracket), ~c \big) \in \mathds{X}$\\
$\mathtt{F} \llbracket ~\Gamma_1 + \ldots + \Gamma_n \Rightarrow \Gamma_{n+1} + \ldots + \Gamma_m~ \rrbracket$ & = &
$(\chi, \omega, \iota, \varphi, \psi) \in \mathds{R}$ such that:\\
\end{tabular}
\end{center}

\begin{center}
\begin{minipage}{0.4\textwidth}
\begin{itemize}
\item $\chi = \big(\mathtt{F} \llbracket ~\Gamma_1~ \rrbracket, \ldots, \mathtt{F} \llbracket ~\Gamma_n~ \rrbracket, \mathtt{F} \llbracket ~\Gamma_{n+1}~ \rrbracket, \ldots, \mathtt{F} \llbracket ~\Gamma_m~ \rrbracket \big)$,
\item $\omega = \mdoubleplus_{i=1}^{|\chi|} \mu(\chi_i)$,
\item $\iota = n$,
\item $\varphi = (J_1, \ldots, J_m)$ ~where~ $J_k = \sum\limits_{i=1}^{k} | \mu(\chi_i) |$,
\end{itemize}
\end{minipage}
\hfill
\begin{minipage}{0.55\textwidth}
\begin{itemize}
\item \begin{tabular}{l l}
& \hspace*{-0.3cm} $\{~ (i,j) ~|~ i \in [1, \varphi_\iota] \wedge j \in [\varphi_\iota + 1, |\omega|] \wedge |i-j| = \varphi_\iota~\} ~\cup$  \\

\hspace*{-0.3cm}$\psi =$ & \hspace*{-0.3cm} $\{~ (i, -) ~|~ i \in [k, \varphi_\iota] \wedge k = | \omega | - \varphi_\iota + 1 ~| ~\} ~\cup$  \\

& \hspace*{-0.3cm} $ \{~ (-, j) ~|~ j \in [k, |\omega|] \wedge k = 2 \times \varphi_\iota + 1 ~\}$
\end{tabular}

\noindent where $\psi$ is defined together with an ordering such that symbol $'-' > k$ for every $k \in \mathbb{N}$ and all descending intervals in definition of $\psi$ are ignored.

\end{itemize}
\end{minipage}
\end{center}
\end{definition}

Note that the translation function works \emph{only} on expressions defined in Definition~\ref{grammar}. The function recursively creates objects from given expressions. Every rule expression is first decomposed to LHS and RHS, and consequently each agent expression is translated to an object. The appropriate index maps are created from sequence of complexes $\chi$ and sequence of atomic and structure agents $\omega$.

\subsection{BCSL model}
\label{BCSl_model}

We proceed to the BCSL model definition. We always consider an initialised model, which means the definition contains an initial state of the system (a solution, Definition~\ref{solution}). The definition of BCSL model also contains rules and signatures.

\begin{definition}{(\emph{Solution})}
\label{solution}
\emph{Solution} is a multiset $\mathcal{S} \in \Omega^\mathds{X}$ such that $\mathcal{A}$ is the reference set of $\mathcal{S}$ and $\forall \mathtt{X} \in \mathcal{A}: \triangle \mathtt{X}$.
\end{definition}

\begin{definition}{(\emph{BCSL model})}
\label{BCSL_model}
We define \emph{BCSL model} $\mathcal{M}$ as a quadruple $(\mathcal{R}, \Sigma_\mathtt{A}, \Sigma_\mathtt{T}, \mathcal{S})$ where $\mathcal{R}$ is a set of rules, $\Sigma_\mathtt{A}$ is an atomic signature, $\Sigma_\mathtt{T}$ is a structure signature, and $\mathcal{S}$ is an initial solution.
\end{definition}

A BCSL model is formed by a set of rules $\mathcal{R}$, which define the behaviour of the model. The initial solution $\mathcal{S}$ defines the state of the model in the beginning. Atomic signature $\Sigma_\mathtt{A}$ defines allowed states for all atomic agents used in the rules. Finally, structure signature $\Sigma_\mathtt{T}$ defines allowed atomic agents for all structure agents used in the rules.

\subsection{Matching}

At this point, we define matching, which will be used in the definition of semantics for a BCSL model $\mathcal{M}$.

\begin{definition}{(\emph{Matching})}
\label{matching}
Let $\mathtt{R} = (\chi, \omega, \iota, \varphi, \psi)$, $\mathtt{r} = (\chi', \omega', \iota', \varphi', \psi')$ be two rules, $\mathcal{S} \in \Omega^\mathds{X}$ be a solution, and $i, j \in \mathbb{N}$.
Let $\models\hspace{3pt}\subseteq \mathds{R} \times \Omega^\mathds{X} \times \mathds{R}$ be the \emph{matching} relation such that a tuple $(\mathtt{R}, \mathcal{S}, \mathtt{r}) \in\hspace{3pt}\models$, written $\mathtt{R} \models_\mathtt{r} \mathcal{S}$, \emph{iff}

\vspace{10pt}

\begin{minipage}{0.35\textwidth}
\begin{enumerate}
  \item $ \iota = \iota' \wedge \varphi = \varphi' \wedge \psi = \psi'$,
  \item $|\chi| = |\chi'| \wedge |\omega| = |\omega'|$,
  \item $\forall i \in [1, |\chi|]: \chi'_i \lhd \chi_i$,
  \item $\Omega(\mathit{LHS}(\mathtt{r})) = \mathcal{S}$,
\end{enumerate}
\end{minipage}
\hfill
\begin{minipage}{0.65\textwidth}
\begin{enumerate}
  \setcounter{enumi}{4}
  \item $\forall (i,j) \in \psi:$
  \begin{enumerate}
    \item $\omega_i \in \mathds{A} \Rightarrow 
      \begin{cases}
        \omega'_i = \omega'_j & \emph{if}~~~~ \omega_i = \omega_j \\
        \omega_i = \omega'_i \wedge \omega_j = \omega'_j & \emph{if}~~~~ \omega_i \neq \omega_j\\
      \end{cases}
          $
    \item $\omega_i \in \mathds{T} \Rightarrow \gamma(\omega'_i) \ominus \gamma(\omega_i) = \gamma(\omega'_j) \ominus \gamma(\omega_j)$.
  \end{enumerate}
\end{enumerate}
\end{minipage}

\vspace{10pt}

\end{definition}

\begin{remark}
Note the rule $\mathtt{r}$ from the tuple $(\mathtt{R}, \mathcal{S}, \mathtt{r}) \in\hspace{4pt}\models$ is so-called \emph{reaction}, which is characterised as an~instance of the rule $\mathtt{R}$. For every rule in a model, it is possible to enumerate all potential reactions and this way convert a rule-based model to a reaction-based model.
\end{remark}

\subsection{Semantics}

\begin{definition}{(\emph{Replacement})}
Let $\rightarrow\hspace{3pt}\subseteq \Omega^\mathds{X} \times \mathds{R} \times \Omega^\mathds{X}$ be the \emph{replacement} relation such that a tuple $(\mathcal{S}, \mathtt{R}, \mathcal{S'}) \in\hspace{3pt}\rightarrow$, written $\mathcal{S} \rightarrow_{\mathtt{R}} \mathcal{S'}$, \emph{iff} $\exists \mathtt{r} \in \mathds{R}$ $\exists x \subseteq \mathcal{S}$ such that 
$\mathtt{R} \models_\mathtt{r} x \wedge~ \mathcal{S'} \setminus (\mathcal{S} \setminus x) = \Omega(\mathit{RHS}(\mathtt{r}))$.
\end{definition}

Replacement relation defines how a solution is transformed according to a given rule. For a BCSL model $\mathcal{M}$, rules yield a labelled transition system $LTS(\mathcal{M})$ between solutions containing an edge $\mathcal{S} \rightarrow_{\mathtt{R}} \mathcal{S'}$. Note that we can achieve the equivalent behaviour if we first generate all possible reactions from the rules and apply replacement with them instead (a rule is just a generalised set of reactions).

\section{Syntactic extensions}
\label{syntactic_extensions}

In this section, we define several syntactic extensions which increase the readability of the rule expressions. Note that each rule expression in an extended form can always be translated to basic form defined above (Section~\ref{syntax}). All rule expressions containing the following extensions must be converted to basic form before the semantics can be applied.
For better demonstration, we provide a running example, which will go through all syntactic extensions (Running example~\ref{run1}). Please note there is no biological sense of the example model, its only purpose is to effectively demonstrate all defined syntactic extensions.

\begin{runningExample}{(\emph{The example model $\mathcal{M}$})}
\label{run1}

{\small
\begin{enumerate}
\item $KaiC(S\{u\}, T\{\varepsilon\}).KaiC(S\{\varepsilon\}, T\{\varepsilon\}).KaiC(S\{\varepsilon\}, T\{\varepsilon\})::cyt \Rightarrow  KaiC(S\{p\}, T\{\varepsilon\}).KaiC(S\{\varepsilon\}, T\{\varepsilon\}).KaiC(S\{\varepsilon\}, T\{\varepsilon\})::cyt$

\item $KaiC(S\{u\}, T\{\varepsilon\}).KaiB(\varnothing)::cyt \Rightarrow KaiC(S\{p\}, T\{\varepsilon\}).KaiB(\varnothing)::cyt$

\item $KaiC(S\{\varepsilon\}, T\{\varepsilon\})::cyt + KaiC(S\{\varepsilon\}, T\{\varepsilon\})::cyt + KaiC(S\{\varepsilon\}, T\{\varepsilon\})::cyt \Rightarrow KaiC(S\{\varepsilon\}, T\{\varepsilon\}).KaiC(S\{\varepsilon\}, T\{\varepsilon\}).KaiC(S\{\varepsilon\}, T\{\varepsilon\})::cyt$

\item $KaiC(S\{\varepsilon\}, T\{\varepsilon\}).KaiC(S\{\varepsilon\}, T\{\varepsilon\}).KaiC(S\{\varepsilon\}, T\{\varepsilon\})::cyt \Rightarrow KaiC(S\{\varepsilon\}, T\{\varepsilon\})::cyt + KaiC(S\{\varepsilon\}, T\{\varepsilon\})::cyt + KaiC(S\{\varepsilon\}, T\{\varepsilon\})::cyt$
\end{enumerate}

\begin{center}
\begin{tabular}{l}
$\Sigma_\mathtt{A} = \big\{~ S \rightarrow \{ u, p \}, T \rightarrow \{ a, i \} ~\big\}$ \\
$\Sigma_\mathtt{T} = \big\{~ KaiC \rightarrow \{ S, T \}, KaiB \rightarrow \emptyset~\big\}$
\end{tabular}
\end{center}
}

\noindent We omit the initial state definition just for simplicity of the example since all the extensions concern only rule expressions.
\end{runningExample}

\subsection{Partial composition context elimination}

It is possible to omit all atomic expressions with unspecified state $\varepsilon$ from partial compositions of structure agents (Running example~\ref{run2}). Such agent expressions do not give any additional information and whole partial composition can be reconstructed from the given signature.

\newpage
\begin{runningExample}{(\emph{The example model $\mathcal{M}$})}
\label{run2}
\begin{enumerate}
\item $KaiC(S\{u\}).KaiC(\varnothing).KaiC(\varnothing)::cyt \Rightarrow KaiC(S\{p\}).KaiC(\varnothing).KaiC(\varnothing)::cyt$
\item $KaiC(S\{u\}).KaiB(\varnothing)::cyt \Rightarrow KaiC(S\{p\}).KaiB(\varnothing)::cyt$
\item $KaiC(\varnothing)::cyt + KaiC(\varnothing)::cyt + KaiC(\varnothing)::cyt \Rightarrow KaiC(\varnothing).KaiC(\varnothing).KaiC(\varnothing)::cyt$
\item $KaiC(\varnothing).KaiC(\varnothing).KaiC(\varnothing)::cyt \Rightarrow KaiC(\varnothing)::cyt + KaiC(\varnothing)::cyt + KaiC(\varnothing)::cyt$
\end{enumerate}
\end{runningExample}

Additionally, this extension can go even further by omitting the $(\varnothing)$ part from structure agents completely (Running example~\ref{run3}). Since we have the structure signature $\Sigma_\mathtt{T}$ defined, we can unambiguously determine which names belong to structure agents and this syntactic part can be easily reconstructed.

\begin{runningExample}{(\emph{The example model $\mathcal{M}$})}
\label{run3}

\begin{enumerate}
\item $KaiC(S\{u\}).KaiC.KaiC::cyt \Rightarrow KaiC(S\{p\}).KaiC.KaiC::cyt$
\item $KaiC(S\{u\}).KaiB::cyt \Rightarrow KaiC(S\{p\}).KaiB::cyt$
\item $KaiC::cyt + KaiC::cyt + KaiC::cyt \Rightarrow KaiC.KaiC.KaiC::cyt$
\item $KaiC.KaiC.KaiC::cyt \Rightarrow KaiC::cyt + KaiC::cyt + KaiC::cyt$
\end{enumerate}
\end{runningExample}

This syntactic extension brings a lot of readability to the syntax while preserving all information in the context of the model $\mathcal{M}$.

\subsection{Complex signature}
\label{complex_signature}

We extend the model definition by complex signature $\Sigma_\mathtt{X}$ (Running example~\ref{run4}). In this signature, there are defined aliases for valid complex expressions. Then, the original complex expressions are substituted by the aliases.

\begin{runningExample}{(\emph{The example model $\mathcal{M}$})}
\label{run4}

\noindent Definition of complex signature $\Sigma_\mathtt{X}$ = $\Set{\begin{array}{l}
KaiC3::cyt \rightarrow KaiC.KaiC.KaiC::cyt,\\
KaiBC::cyt \rightarrow KaiC.KaiB::cyt
\end{array}}$

\begin{enumerate}
\item $KaiC(S\{u\}).KaiC.KaiC::cyt \Rightarrow KaiC(S\{p\}).KaiC.KaiC::cyt$
\item $KaiC(S\{u\}).KaiB::cyt \Rightarrow KaiC(S\{p\}).KaiB::cyt$
\item $KaiC::cyt + KaiC::cyt + KaiC::cyt \Rightarrow KaiC3::cyt$
\item $KaiC3::cyt \Rightarrow KaiC::cyt + KaiC::cyt + KaiC::cyt$
\end{enumerate}
\end{runningExample}

The usage of the complex signature has its limitations. Once a context is specified, the alias cannot be used. We will resolve this problem in the following extensions.

\subsection{Directions}

We allow rule expressions to be bi-directional -- it is just a shortcut for two rule expressions and it can be converted to the basic rule expression form. A rule expression $\varrho : l ~\Leftrightarrow~ r$ can be written as two rule expressions $\varrho_1 : l ~\Rightarrow~ r$ and $\varrho_2 : r ~\Rightarrow~ l$ (Running example~\ref{run5}).

\begin{runningExample}{(\emph{The example model $\mathcal{M}$})}
\label{run5}

\begin{enumerate}
\item $KaiC(S\{u\}).KaiC.KaiC::cyt \Rightarrow KaiC(S\{p\}).KaiC.KaiC::cyt$
\item $KaiC(S\{u\}).KaiB::cyt \Rightarrow KaiC(S\{p\}).KaiB::cyt$
\item $KaiC::cyt + KaiC::cyt + KaiC::cyt \Leftrightarrow KaiC3::cyt$
\end{enumerate}

\end{runningExample}

\noindent Definition of rules (iii) and (iv) from Running example~\ref{run4} was replaced by one bi-directional rule (iii) in Running example~\ref{run5}.

\subsection{Stoichiometry}

For a rule expression of form:

\begin{center}
$\beta_1::\mathtt{c} + \beta_2::\mathtt{c} + \ldots + \beta_n::\mathtt{c} \Rightarrow \beta_1.\beta_2.~\ldots~.\beta_n::\mathtt{c}$
\end{center}

\noindent we can reorder both sides such that we get non-crossing partition $\mathcal{P} = B_1/B_2/\ldots/B_k$ with $k \leq n$ from its indices $[1,\ldots,n]$ such that:
$\forall B \in \mathcal{P} ~\forall \beta, \beta' \in B: \beta = \beta'$ 
and
$\forall B, B' \in \mathcal{P} ~\forall \beta \in B ~\forall \beta' \in B': \beta \neq \beta'$
such that $B \neq B'$.

For the left-hand side $\beta_1::\mathtt{c} + \beta_2::\mathtt{c} + \ldots + \beta_n::\mathtt{c}$ of the reordered rule expression we can replace all rule expressions $[\beta_i, \ldots, \beta_j]$ which belong to the same non-crossing partition $B$ by notation $`k~\beta'$, where $\beta$ is a representative from $\beta_i, \ldots, \beta_j$ (they are all equivalent) and $k$ is the number of the expressions in partition $B$ (Running example~\ref{run6}). Note that this process is fully reversible -- we can simply enumerate all expressions for each partition.

\begin{runningExample}{(\emph{The example model $\mathcal{M}$})}
\label{run6}

\noindent Definition of rule expressions:
\begin{enumerate}
\item $KaiC(S\{u\}).KaiC.KaiC::cyt \Rightarrow KaiC(S\{p\}).KaiC.KaiC::cyt$
\item $KaiC(S\{u\}).KaiB::cyt \Rightarrow KaiC(S\{p\}).KaiB::cyt$
\item $3~KaiC::cyt \Leftrightarrow KaiC3::cyt$
\end{enumerate}
\end{runningExample}

\noindent Definition of rule expression (iii) from Running example~\ref{run5} was replaced by a new rule expression using stoichiometry.

\subsection{Locations}

The localisation operator is intended for allowing an alternative way of expressing the hierarchically constructed agent expressions (Running example~\ref{run7}). The main idea is to allow zooming into individual parts of complex and structure expressions. For this purpose, we use $a :: b$ notation such that $a, b$ are arbitrary agents which satisfy one of the conditions given in Definition~\ref{locations:conditions}.

\begin{definition}{(\emph{Location conditions})}\label{locations:conditions}

\begin{enumerate}
 \item $\mathtt{A}::\mathtt{T}$ $\Leftrightarrow$ there exists $\mathtt{A}' \in \gamma(\mathtt{T})$ such that $\mathtt{A} \lhd \mathtt{A}'$,

\item $\mathtt{A}::\mathtt{X}$ $\Leftrightarrow$ there exists $\mathtt{A}' \in \mu(\mathtt{X})$ such that $\mathtt{A} \lhd \mathtt{A}'$,

\item $\mathtt{T}::\mathtt{X}$ $\Leftrightarrow$ there exists $\mathtt{T}' \in \mu(\mathtt{X})$ such that $\mathtt{T} \lhd \mathtt{T}'$.
\end{enumerate}
\end{definition}

For each pair of agents $(\alpha, \beta)$ with allowed `::' operator between them, we can construct just one agent $\beta'$ without the operator by taking the most left agent $\alpha'$ from full (resp. partial) composition of the agent $\beta$ such that it is \emph{compatible with} the agent $\alpha$. Then, agent $\alpha'$ is merged with agent $\alpha$ and agent $\beta'$ is constructed.

\begin{runningExample}{(\emph{The example model $\mathcal{M}$})}
\label{run7}

\begin{enumerate}
\item $S\{u\}::KaiC::KaiC3::cyt \Rightarrow S\{p\}::KaiC::KaiC3::cyt$
\item $S\{u\}::KaiC::KaiBC::cyt \Rightarrow S\{p\}::KaiC::KaiBC::cyt$
\item $3~KaiC::cyt \Leftrightarrow KaiC3::cyt$
\end{enumerate}
\end{runningExample}

Definition of rule expressions (i) and (ii) from Running example~\ref{run6} was replaced using locations. The localisation operator allowed us to additionally use the complex signatures.

\subsection{Variables}

Rule expressions (i) and (ii) from Running example~\ref{run7} are very similar except for the context of complex expression they take place in. We can substitute this context with a variable with a given domain.

In a rule expression, one agent expression might be referenced using a variable as a set of rule agent expressions it can be replaced with (Running example~\ref{run8}). Such an agent expression is referenced as $?X$. Moreover, in the case when a $?X$ is used in a location, it must hold conditions from Definition~\ref{locations:conditions}.

Each rule expression associated with a variable can be easily written as several rule expressions where the variable is replaced with agent expression from the set of agent expressions attached to the variable. For simplicity, only one variable can be used per rule expression.

\begin{runningExample}{(\emph{The example model $\mathcal{M}$})}
\label{run8}

\begin{enumerate}
\item $S\{u\}::KaiC::~?X::cyt \Rightarrow S\{p\}::KaiC::~?X::cyt ~;~ ?X = \{KaiC3, KaiBC\}$
\item $3~KaiC::cyt \Leftrightarrow KaiC3::cyt$
\end{enumerate}
\end{runningExample}

\noindent Definition of rule expressions (i) and (ii) from Running example~\ref{run7} was replaced as a single rule expression with a variable.

This is the final syntactic extension. Compared to the original model (Running example~\ref{run1}), the resulting model is more concise and readable.

\section{Static analysis}
\label{static_analyses}

The BCS language offers interesting capabilities to provide several static analysis techniques of given models. These techniques are based on defined \emph{compatibility} operator $\lhd$, which formulates suitable properties for each type of agent.

\begin{definition}{(\emph{Ordering of agents})}
\label{partial_order}
Let $x_1, x_2$ be two arbitrary agents. The compatibility relation induces \emph{partial ordering} of agents $x_1$ and $x_2$, written $x_1 \leq x_2$, \emph{iff} $ x_1 \lhd x_2 $.
\end{definition}

\begin{notation}
The universe of complex agents $\mathds{X}$ with partial order $\leq$ is a partially ordered set $\mathds{X}_\leq$.
\end{notation}

The compatibility operator defines a partial order on $\mathds{A}, \mathds{T}$, and $\mathds{X}$ sets. For our purposes, only partially ordered set $\mathds{X}_\leq$ is relevant. The reason is that complex agents actually encapsulate all the other agent types. However, partial order of the entire universe of complex agents is not very useful, since most of the agents cannot be compared by compatibility operator. We are interested in particular subsets where every two complex agents can be either compared directly or there exists an agent compatible with both of them.

\begin{definition}{(\emph{Compatible set})}
\label{compatible_set}
A finite set $\mathcal{X} \subseteq \mathds{X}$ is a \emph{compatible set} if:
\begin{center}
\begin{enumerate}
	\item \label{compatible_cond} $ \forall \mathtt{X}_1, \mathtt{X}_2 \in \mathcal{X} ~\exists \mathtt{X}' \in \mathcal{X}: \mathtt{X}_1 \lhd \mathtt{X}' \wedge \mathtt{X}_2 \lhd \mathtt{X}'$,

  \item \label{non_compatible_cond} and for each finite set $\mathcal{X}' \subseteq \mathds{X}$ such that $\mathcal{X} \cap \mathcal{X}' = \emptyset$ holds:
	$\forall \mathtt{X} \in \mathcal{X} ~\forall \mathtt{X}' \in \mathcal{X}': \neg (\mathtt{X} \lhd \mathtt{X}' \vee \mathtt{X}' \lhd \mathtt{X}) $.
\end{enumerate}
\end{center}

\end{definition}

\begin{remark}
\label{order_in_compat_set}
The compatible set $\mathcal{X}$ inherits partial order of $\mathds{X}_\leq$ since it is its subset.
\end{remark}

\begin{figure}[!h]
\begin{center}
\includegraphics[scale=0.15]{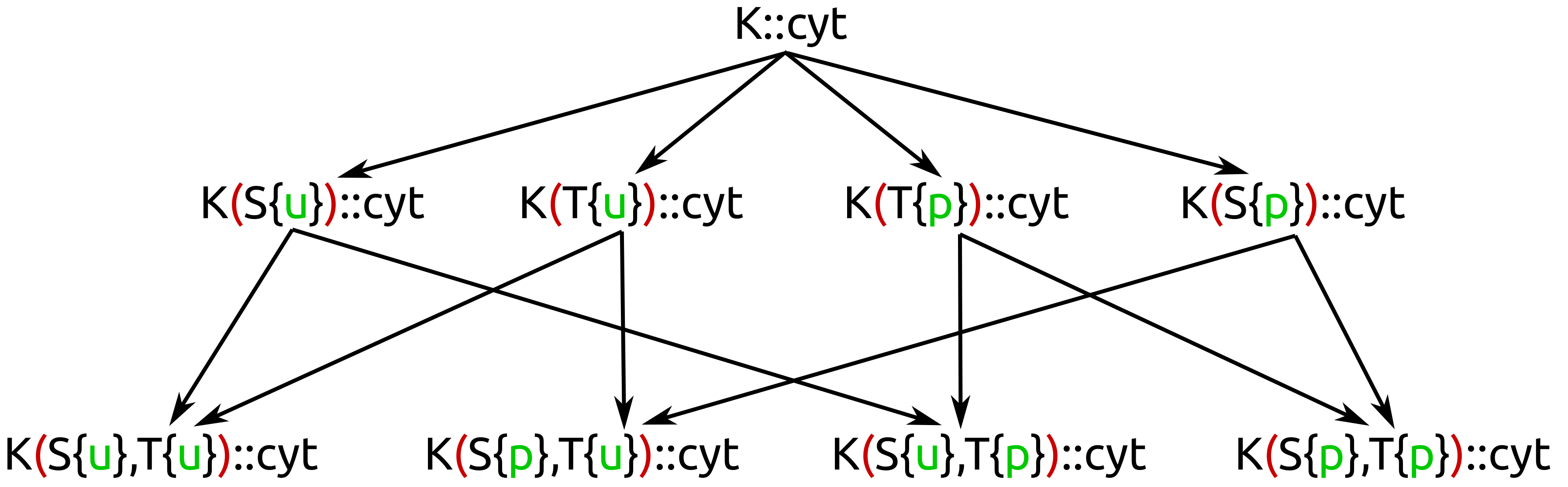}
\end{center}
\caption{An example of a compatible set $\mathcal{X}$. The set is formed by a complex in $cyt$ compartment, which has only one structure agent $K$ in its sequence. The structure agent $K$ has allowed atomic agents $T$ and $S$ in its partial composition. These two atomic agents might occur in two states -- $\textcolor{mydarkgreen}{u}$ and $\textcolor{mydarkgreen}{p}$. The set is complete -- there are all relevant agents bounded by compatibility operator.}\label{compatible_set_example}
\end{figure}

A compatible set $\mathcal{X}$ contains partially ordered complex agents such that they all have the same sequences in terms of agent names. Example of a compatible set is given in Figure~\ref{compatible_set_example}.

\begin{lemma}
\label{global_supremum}
In every compatible set $\mathcal{X}$, there always exists a global supremum $\mathtt{sup}(\mathcal{X})$.

\begin{proof}
The lemma follows from Definition~\ref{compatible_set} condition~(\ref{compatible_cond}) which claims that there is a supremum (in terms of compatibility) for every two complex agents in the compatible set $\mathcal{X}$. Since there exists a supremum for every two items in the set and the set is finite, there must exist a global supremum for the entire set. 
\end{proof}
\end{lemma}

\begin{lemma}
\label{exists_one_comp_set}
For every complex agent $\mathtt{X}$ there exists exactly one compatible set $\mathcal{X} \subseteq \mathds{X}$ such that $ \mathtt{X} \in \mathcal{X}$.

\begin{proof}
Let us assume a complex agent $\mathtt{X}$ belongs to two compatible sets, namely $\mathtt{X} \in \mathcal{X}_1, \mathcal{X}_2$. From Definition~\ref{compatible_set} condition~(\ref{compatible_cond}) follows that there exists a $\mathtt{X}_1 \in \mathcal{X}_1$ such that $\mathtt{X} \lhd \mathtt{X}_1$. 

Next, the condition~(\ref{non_compatible_cond}) claims that no complex agent from $\mathcal{X}_1$ and no complex agent from $\mathcal{X}_2$ can be compatible. Namely, $\mathtt{X}_1 \in \mathcal{X}_1$ cannot be compatible with $\mathtt{X} \in \mathcal{X}_2$. However, $\mathtt{X}$ and $\mathtt{X}_1$ are compatible ($\mathtt{X} \lhd \mathtt{X}_1$). It follows $\mathtt{X} \not\in \mathcal{X}_2$, which is a contradiction. 
\end{proof}
\end{lemma}

In practise, compatible sets can be used for finding non-trivial relationships between the rules (Section~\ref{rule_redundancy}) and for static analysis on the level of complexes (Section~\ref{context_reduction}).

\begin{definition}{(\emph{Compatible subset})}
\label{compatible_subset}
Let $\mathcal{X} \subseteq \mathds{X}$ be a compatible set and $\mathtt{X} \in \mathcal{X}$ a complex agent. A set $\mkern2mu\overline{\mkern-4.5mu \mathcal{X}} \subseteq \mathcal{X}$ is called \emph{compatible subset} of $\mathcal{X}$ w.r.t. $\mathtt{X}$ if the following conditions hold:
\begin{enumerate}
\item $ \forall \mathtt{X}' \in \mkern2mu\overline{\mkern-4.5mu \mathcal{X}}: \mathtt{X}' \lhd \mathtt{X} \wedge \triangle \mathtt{X}'$,
\item $\not\exists \mathtt{X}'' \in \mathcal{X} \setminus \mkern2mu\overline{\mkern-4.5mu \mathcal{X}}: \mathtt{X}'' \lhd \mathtt{X} \wedge \triangle \mathtt{X}''$.
\end{enumerate}
\end{definition}

Compatible subset formally defines all fully specified agents from the compatible set which are compatible with a given member of the set (i.e. there are no compatible agents with them in the set). Note that for any complex agent $\mathtt{X}$ there exists just one compatible subset. The reason follows from Lemma~\ref{exists_one_comp_set} and Definition~\ref{compatible_subset}.

\subsection{Rule redundancy elimination}
\label{rule_redundancy}

There might be cases where there are redundant rules in a model (Definition~\ref{redundant_rule}). These rules do not cause any semantic difference, only increase the size of the model. We provide a static method how to detect such rules and eventually delete them from the model. Please note the redundancy is relevant only in the qualitative context. In the quantitative context, the same rules with different kinetics might have their relevance, yet it is still useful to detect potential redundancies.

\begin{definition}{(\emph{Redundant rule})}
\label{redundant_rule}
Let $\mathcal{M}_1 = (\mathcal{R} \cup \{ \mathtt{R} \}, \Sigma_\mathtt{A}, \Sigma_\mathtt{T}, \mathcal{S})$ and
$\mathcal{M}_2 = (\mathcal{R}, \Sigma_\mathtt{A}, \Sigma_\mathtt{T}, \mathcal{S})$ be BCSL models where $\mathtt{R}$ is a rule such that $\mathtt{R} \not\in \mathcal{R}$. The rule $\mathtt{R}$ is \emph{redundant} if $LTS(\mathcal{M}_1) = LTS(\mathcal{M}_2)$.
\end{definition}

The redundant rule $\mathtt{R}$ does not add any semantic information to the model. It generally means the LTSs produced from the models with and without the rule are equal.

\begin{theorem}
\label{redundant_if}
Let $\mathtt{R} = (\chi, \omega, \iota, \varphi, \psi)$ and $\mathtt{R}' = (\chi', \omega', \iota', \varphi', \psi')$ be two rules such that $|\chi| = |\chi'| = n$ for some $n \in \mathbb{N}$. The rule $\mathtt{R}'$ is \emph{redundant} if $ \forall i \in [ 1, n ]: \chi'_i \lhd \chi_i$.

\begin{proof}
The problem whether the elimination of a redundant rule preserves semantics can be reduced to a simple question -- if it holds for a single pair of complex agents for a position \emph{k} in the appropriate rules, then it generally holds for entire rule, because the condition of redundancy holds for each pair of complexes independently.

Assume the complex agents $\mathtt{X}_k$ and $\mathtt{X}'_k$ both belong to the same compatible set $\mathcal{X}$ since $\mathtt{X}_k \lhd \mathtt{X}'_k$, which follows from the condition of the theorem. 
We can create subsets ~$\mkern2mu\overline{\mkern-4.5mu \mathcal{X}},~ \mkern2mu\overline{\mkern-4.5mu \mathcal{X}'} \subseteq \mathcal{X}$ for both complex agents respectively (Definition~\ref{compatible_subset}). Since the agents are compatible ($\mathtt{X}_k \lhd \mathtt{X}'_k$), the compatible subset ~$\mkern2mu\overline{\mkern-4.5mu \mathcal{X}}$ w.r.t. agent $\mathtt{X}_k$ is subset of the compatible subset ~$\mkern2mu\overline{\mkern-4.5mu \mathcal{X}'}$ w.r.t. agent $\mathtt{X}'_k$ (~$\mkern2mu\overline{\mkern-4.5mu \mathcal{X}} \subseteq ~\mkern2mu\overline{\mkern-4.5mu \mathcal{X}'}$).

Applied generally on the entire rule, the produced set of reactions (using matching relation -- Definition~\ref{matching}) from the redundant rule is actually a subset of reactions produced from the non-redundant rule.
\end{proof}
\end{theorem}

In the proof, we used compatible sets of complex agents and the fact that we can generate reactions from the rules, while we are actually enumerating all agents from the compatible set which are \emph{compatible with} original agent in the rule. This is demonstrated in Example~\ref{redundant_example}.

\begin{example}{Redundant rule.}
\label{redundant_example}
Let us consider two rules:

\begin{center}
\begin{minipage}{0.5\textwidth}
\begin{enumerate}
  \item \label{rule_red_1} $K\textcolor{myred}{(}S\{\textcolor{mydarkgreen}{u}\}\textcolor{myred}{)}.K::cell \Rightarrow K\textcolor{myred}{(}S\{\textcolor{mydarkgreen}{p}\}\textcolor{myred}{)}.K::cell$
  \item \label{rule_red_2} $K\textcolor{myred}{(}S\{\textcolor{mydarkgreen}{u}\}, T\{\textcolor{mydarkgreen}{i}\}\textcolor{myred}{)}.K::cell \Rightarrow K\textcolor{myred}{(}S\{\textcolor{mydarkgreen}{p}\}, T\{\textcolor{mydarkgreen}{i}\}\textcolor{myred}{)}.K::cell$
\end{enumerate}
\end{minipage}
\end{center}

Considering structure signature $\Sigma_\mathtt{T}(K) = \{ S, T \} $ and atomic signatures $\Sigma_\mathtt{A}(S) = \{ u, p \} $ and $\Sigma_\mathtt{A}(T) = \{ a, i \} $, the~rule (\ref{rule_red_1}) produces following set of eight reactions:

\begin{center}
$\Set{\begin{array}{l}
K\textcolor{myred}{(}S\{\textcolor{mydarkgreen}{u}\}, T\{\textcolor{mydarkgreen}{a}\}\textcolor{myred}{)}.K\textcolor{myred}{(}S\{\textcolor{mydarkgreen}{u}\}, T\{\textcolor{mydarkgreen}{a}\}\textcolor{myred}{)}::cell \Rightarrow K\textcolor{myred}{(}S\{\textcolor{mydarkgreen}{p}\}, T\{\textcolor{mydarkgreen}{a}\}\textcolor{myred}{)}.K\textcolor{myred}{(}S\{\textcolor{mydarkgreen}{u}\}, T\{\textcolor{mydarkgreen}{a}\}\textcolor{myred}{)}::cell,\\
K\textcolor{myred}{(}S\{\textcolor{mydarkgreen}{u}\}, T\{\textcolor{mydarkgreen}{a}\}\textcolor{myred}{)}.K\textcolor{myred}{(}S\{\textcolor{mydarkgreen}{u}\}, T\{\textcolor{mydarkgreen}{i}\}\textcolor{myred}{)}::cell \Rightarrow K\textcolor{myred}{(}S\{\textcolor{mydarkgreen}{p}\}, T\{\textcolor{mydarkgreen}{a}\}\textcolor{myred}{)}.K\textcolor{myred}{(}S\{\textcolor{mydarkgreen}{u}\}, T\{\textcolor{mydarkgreen}{i}\}\textcolor{myred}{)}::cell,\\
K\textcolor{myred}{(}S\{\textcolor{mydarkgreen}{u}\}, T\{\textcolor{mydarkgreen}{a}\}\textcolor{myred}{)}.K\textcolor{myred}{(}S\{\textcolor{mydarkgreen}{p}\}, T\{\textcolor{mydarkgreen}{a}\}\textcolor{myred}{)}::cell \Rightarrow K\textcolor{myred}{(}S\{\textcolor{mydarkgreen}{p}\}, T\{\textcolor{mydarkgreen}{a}\}\textcolor{myred}{)}.K\textcolor{myred}{(}S\{\textcolor{mydarkgreen}{p}\}, T\{\textcolor{mydarkgreen}{a}\}\textcolor{myred}{)}::cell,\\
K\textcolor{myred}{(}S\{\textcolor{mydarkgreen}{u}\}, T\{\textcolor{mydarkgreen}{a}\}\textcolor{myred}{)}.K\textcolor{myred}{(}S\{\textcolor{mydarkgreen}{p}\}, T\{\textcolor{mydarkgreen}{i}\}\textcolor{myred}{)}::cell \Rightarrow K\textcolor{myred}{(}S\{\textcolor{mydarkgreen}{p}\}, T\{\textcolor{mydarkgreen}{a}\}\textcolor{myred}{)}.K\textcolor{myred}{(}S\{\textcolor{mydarkgreen}{p}\}, T\{\textcolor{mydarkgreen}{i}\}\textcolor{myred}{)}::cell,\\
K\textcolor{myred}{(}S\{\textcolor{mydarkgreen}{u}\}, T\{\textcolor{mydarkgreen}{i}\}\textcolor{myred}{)}.K\textcolor{myred}{(}S\{\textcolor{mydarkgreen}{u}\}, T\{\textcolor{mydarkgreen}{a}\}\textcolor{myred}{)}::cell \Rightarrow K\textcolor{myred}{(}S\{\textcolor{mydarkgreen}{p}\}, T\{\textcolor{mydarkgreen}{i}\}\textcolor{myred}{)}.K\textcolor{myred}{(}S\{\textcolor{mydarkgreen}{u}\}, T\{\textcolor{mydarkgreen}{a}\}\textcolor{myred}{)}::cell,\\
K\textcolor{myred}{(}S\{\textcolor{mydarkgreen}{u}\}, T\{\textcolor{mydarkgreen}{i}\}\textcolor{myred}{)}.K\textcolor{myred}{(}S\{\textcolor{mydarkgreen}{u}\}, T\{\textcolor{mydarkgreen}{i}\}\textcolor{myred}{)}::cell \Rightarrow K\textcolor{myred}{(}S\{\textcolor{mydarkgreen}{p}\}, T\{\textcolor{mydarkgreen}{i}\}\textcolor{myred}{)}.K\textcolor{myred}{(}S\{\textcolor{mydarkgreen}{u}\}, T\{\textcolor{mydarkgreen}{i}\}\textcolor{myred}{)}::cell,\\
K\textcolor{myred}{(}S\{\textcolor{mydarkgreen}{u}\}, T\{\textcolor{mydarkgreen}{i}\}\textcolor{myred}{)}.K\textcolor{myred}{(}S\{\textcolor{mydarkgreen}{p}\}, T\{\textcolor{mydarkgreen}{a}\}\textcolor{myred}{)}::cell \Rightarrow K\textcolor{myred}{(}S\{\textcolor{mydarkgreen}{p}\}, T\{\textcolor{mydarkgreen}{i}\}\textcolor{myred}{)}.K\textcolor{myred}{(}S\{\textcolor{mydarkgreen}{p}\}, T\{\textcolor{mydarkgreen}{a}\}\textcolor{myred}{)}::cell,\\
K\textcolor{myred}{(}S\{\textcolor{mydarkgreen}{u}\}, T\{\textcolor{mydarkgreen}{i}\}\textcolor{myred}{)}.K\textcolor{myred}{(}S\{\textcolor{mydarkgreen}{p}\}, T\{\textcolor{mydarkgreen}{i}\}\textcolor{myred}{)}::cell \Rightarrow K\textcolor{myred}{(}S\{\textcolor{mydarkgreen}{p}\}, T\{\textcolor{mydarkgreen}{i}\}\textcolor{myred}{)}.K\textcolor{myred}{(}S\{\textcolor{mydarkgreen}{p}\}, T\{\textcolor{mydarkgreen}{i}\}\textcolor{myred}{)}::cell
\end{array}}$
\end{center}

while the rule (\ref{rule_red_2}) produces set of four reactions:

\begin{center}
$\Set{\begin{array}{l}
K\textcolor{myred}{(}S\{\textcolor{mydarkgreen}{u}\}, T\{\textcolor{mydarkgreen}{i}\}\textcolor{myred}{)}.K\textcolor{myred}{(}S\{\textcolor{mydarkgreen}{u}\}, T\{\textcolor{mydarkgreen}{a}\}\textcolor{myred}{)}::cell \Rightarrow K\textcolor{myred}{(}S\{\textcolor{mydarkgreen}{p}\}, T\{\textcolor{mydarkgreen}{i}\}\textcolor{myred}{)}.K\textcolor{myred}{(}S\{\textcolor{mydarkgreen}{u}\}, T\{\textcolor{mydarkgreen}{a}\}\textcolor{myred}{)}::cell,\\
K\textcolor{myred}{(}S\{\textcolor{mydarkgreen}{u}\}, T\{\textcolor{mydarkgreen}{i}\}\textcolor{myred}{)}.K\textcolor{myred}{(}S\{\textcolor{mydarkgreen}{u}\}, T\{\textcolor{mydarkgreen}{i}\}\textcolor{myred}{)}::cell \Rightarrow K\textcolor{myred}{(}S\{\textcolor{mydarkgreen}{p}\}, T\{\textcolor{mydarkgreen}{i}\}\textcolor{myred}{)}.K\textcolor{myred}{(}S\{\textcolor{mydarkgreen}{u}\}, T\{\textcolor{mydarkgreen}{i}\}\textcolor{myred}{)}::cell,\\
K\textcolor{myred}{(}S\{\textcolor{mydarkgreen}{u}\}, T\{\textcolor{mydarkgreen}{i}\}\textcolor{myred}{)}.K\textcolor{myred}{(}S\{\textcolor{mydarkgreen}{p}\}, T\{\textcolor{mydarkgreen}{a}\}\textcolor{myred}{)}::cell \Rightarrow K\textcolor{myred}{(}S\{\textcolor{mydarkgreen}{p}\}, T\{\textcolor{mydarkgreen}{i}\}\textcolor{myred}{)}.K\textcolor{myred}{(}S\{\textcolor{mydarkgreen}{p}\}, T\{\textcolor{mydarkgreen}{a}\}\textcolor{myred}{)}::cell,\\
K\textcolor{myred}{(}S\{\textcolor{mydarkgreen}{u}\}, T\{\textcolor{mydarkgreen}{i}\}\textcolor{myred}{)}.K\textcolor{myred}{(}S\{\textcolor{mydarkgreen}{p}\}, T\{\textcolor{mydarkgreen}{i}\}\textcolor{myred}{)}::cell \Rightarrow K\textcolor{myred}{(}S\{\textcolor{mydarkgreen}{p}\}, T\{\textcolor{mydarkgreen}{i}\}\textcolor{myred}{)}.K\textcolor{myred}{(}S\{\textcolor{mydarkgreen}{p}\}, T\{\textcolor{mydarkgreen}{i}\}\textcolor{myred}{)}::cell
\end{array}}$
\end{center}

which is a subset of the previous one. It follows the rule (\ref{rule_red_2}) is redundant.

\end{example}

\subsection{Context-based reduction}
\label{context_reduction}

There might be cases when simplifying some details of the given BCSL
model preserves some properties while making the analysis of the model
simpler. This is particularly the case of dynamic analysis, where a
minor change in the model specification can dramatically affect the
behaviour. To address the model simplification, we first define a function that simplifies rules and then define the notion of a reduced model and show what kind of information does it preserve.

\begin{definition}{(\emph{Rule reduction})}
Let $\mathtt{R} = (\chi, \omega, \iota, \varphi, \psi)$ be a rule. We define a reduced rule $\mathtt{R}' = (\chi', \omega', \iota', \varphi', \psi')$ as a function $\theta(\mathtt{R})$ such that $\forall i \in [1, k]: \chi'_i= \mathtt{sup}(\mathcal{X})$ where $\mathcal{X}$ is a compatible set such that $\chi_i \in \mathcal{X}$, length $k = |\chi'| = |\chi|$ (i.e. the number of complex agents in both rules is the same), and $\iota = \iota'$.
\end{definition}

\begin{definition}{(\emph{Reduced model})}
\label{reduced_model}
Let $\mathcal{M} = (\mathcal{R}, \Sigma_\mathtt{A}, \Sigma_\mathtt{T}, \mathcal{S})$ be an initial BCSL model. We define \emph{reduced model} $\widetilde{\mathcal{M}} = (\widetilde{\mathcal{R}}, \Sigma_\mathtt{A}, \Sigma_\mathtt{T}, \mathtt{I})$ such that the following conditions hold:

\begin{enumerate}
	\item for every rule $\mathtt{R} \in \mathcal{R}$, $\theta(\mathtt{R}) \in \widetilde{\mathcal{R}}$ and every rule in the reduced model is the image by $\theta$ of a rule of the initial model;

	\item for every complex agent $\mathtt{X} \in \mathcal{S}$, $\mathtt{sup}(\mathcal{X}) \in \mathtt{I}$ where $\mathcal{X}$ is a compatible set such that $\mathtt{X} \in \mathcal{X}$ and every complex agent in the reduced model is the image by $\mathtt{sup}(\mathcal{X})$ of a complex agent of the initial model.
\end{enumerate}
\end{definition}

\begin{figure}[!h]
\begin{center}
\includegraphics[scale=0.4]{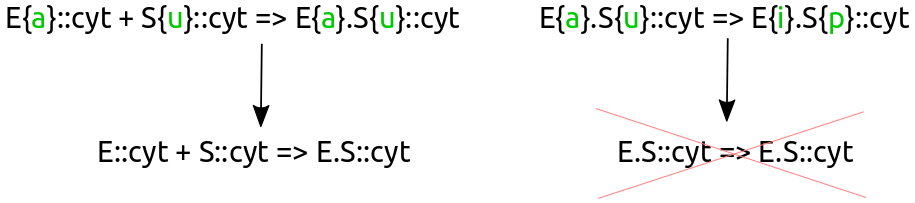}
\end{center}
\caption{Examples of rule reductions. (\emph{left}) A rule of complex formation is reduced to a version where none of the states is specified. (\emph{right}) A rule of state change inside of a complex is reduced to a rule which is not well-formed. It violates the condition~(\ref{cond1}) of Definition~\ref{well_formed} -- an agent has to change during the rule application. Therefore it is removed from the reduced model.}\label{reduction_example}
\end{figure}

Reduced model $\widetilde{\mathcal{M}}$ is created from the given BCSL model by reducing the context of complexes in the rules to the maximum level. This is achieved by taking supremum from compatible set $\mathcal{X}$. This procedure can produce some not well-formed rules -- such rules are omitted (Figure~\ref{reduction_example}). Consequently, only rules creating/destroying agents and complex formation/dissociation should remain. Since we are reducing context, the number of rules in the resulting model is equal to or smaller than the number of rules in the initial model.

\begin{definition}{(\emph{Compatibility of states})}
Let $\mathcal{M}$ be a BCSL model and $\mathtt{s}_1, \mathtt{s}_2$ two states from its LTS. The state $\mathtt{s}_1$ is \emph{compatible with} state $\mathtt{s}_2$, written $\mathtt{s}_1 \lhd \mathtt{s}_2$, \emph{iff} there exists a bijective function $f: \mathtt{s}_1 \rightarrow \mathtt{s}_2$ such that $\forall \mathtt{X} \in \mathtt{s}_1 : \mathtt{sup}(\mathcal{X}) = f(\mathtt{X})$
where $\mathcal{X} \subseteq \mathds{X}$ is a compatible set w.r.t. $\mathtt{X}$.
\end{definition}

\begin{definition}{(\emph{Over-approximation of LTS})}
\label{over_approximation}
Let $LTS(\mathcal{M}), LTS(\mathcal{M}')$ be labelled transition systems of some BCSL models $\mathcal{M}, \mathcal{M}'$. The $LTS(\mathcal{M}')$ is an \emph{over-approximation} of $LTS(\mathcal{M})$ if for every path $\ldots \mathtt{s}_1' \mathtt{s}_2' \mathtt{s}_3' \ldots \mathtt{s}_n' \ldots$ in $LTS(\mathcal{M}')$ there exists a path $\ldots \mathtt{s}_1 \mathtt{s}_2 \mathtt{s}_3 \ldots \mathtt{s}_m \ldots$ in $LTS(\mathcal{M})$ such that
$\forall \mathtt{s}'_i, \mathtt{s}'_{i+1} ~\exists \mathtt{s}_k, \mathtt{s}_l : (l > k \wedge \mathtt{s}_k \lhd \mathtt{s}'_i \wedge \mathtt{s}_l \lhd \mathtt{s}'_{i+1}) $.
\end{definition}

A reduced model $\widetilde{\mathcal{M}}$ is actually an over-approximation of a BCSL model $\mathcal{M}$ in the context of their LTSs (Definition~\ref{over_approximation}). It can be used for some types of analyses which avoid combinatorial explosion of the initial model $\mathcal{M}$.

\begin{theorem}
\label{non-reachability_in_reduced}
Let $\mathtt{X}$ be a complex agent, $\mathcal{X}$ be a compatible set w.r.t. $\mathtt{X}$, $\mathcal{M}$ be a given BCSL model, and $\widetilde{\mathcal{M}}$ be an appropriate reduced model of model $\mathcal{M}$. If supremum $\mathtt{sup}(\mathcal{X})$ is non-reachable in $LTS(\widetilde{\mathcal{M}})$, then agent $\mathtt{X}$ is also non-reachable in the $LTS(\mathcal{M})$.

\begin{proof}
Let us assume a complex agent $\mathtt{sup}(\mathcal{X})$ is non-reachable in $LTS(\widetilde{\mathcal{M}})$, but $\mathtt{X} \in \mathcal{X}$ is reachable in $LTS(\mathcal{M})$. Generally, there is a path formed from rules in the $LTS(\mathcal{M})$ such that we transform complex agents from initial agents to desired complex agent $\mathtt{X}$. When we move to context of $LTS(\widetilde{\mathcal{M}})$, there is no such path for $\mathtt{sup}(\mathcal{X})$.

According to Definition~\ref{reduced_model}, for every such rule there exists a reduced rule, such that all interacting complexes are reduced to their suprema. Therefore, if we could apply an initial rule on a complex agent, we can do the same with reduced rule and its supremum. It follows there must exist such path also in $LTS(\widetilde{\mathcal{M}})$ and the complex agent $\mathtt{sup}(\mathcal{X})$ is reachable, which is a contradiction.
\end{proof}
\end{theorem}

When we are checking whether an agent is reachable in
$LTS(\mathcal{M})$ for given model $\mathcal{M}$, we might first check
whether the respective abstract agent (the supremum) is reachable in
$LTS(\widetilde{\mathcal{M}})$ of the reduced model
$\widetilde{\mathcal{M}}$. If this holds then we are still not certain
about reachability of the agent in its initial form. This has to be checked in $LTS(\mathcal{M})$. However, Theorem~\ref{non-reachability_in_reduced} states that agent which is not reachable in $LTS(\widetilde{\mathcal{M}})$ is also not reachable in $LTS(\mathcal{M})$. The usage of the theorem is demonstrated in Section~\ref{case_study}.

\subsection{Static non-reachability analysis}
\label{static_reachability_analysis}

Since we have defined the compatibility operator for agents, we can apply static non-reachability analysis before enumerating the entire transition system of the model $\mathcal{M}$. We can use the fact that there has to exist a compatible agent on the right-hand side of a rule with the desired agent in order to construct it eventually. This analysis is independent of the initial state of the model. However, it is worth noting that we do not consider the trivial case when the desired agent is already in the initial state.

\begin{theorem}
\label{static_reach}
Let $\mathcal{M}$ be a BCSL model and $\mathcal{R}$ its set of rules. Let $\mathtt{X}$ be a complex agent. The complex agent $\mathtt{X}$ is non-reachable w.r.t. set of rules $\mathcal{R}$ if the following holds:
$\forall~ \mathtt{R} \in \mathcal{R} ~\forall i \in \mathit{RHS}(\mathtt{R}) : \neg ( \chi_i \lhd \mathtt{X})$,
where $\mathtt{R} = (\chi, \omega, \iota, \varphi, \psi)$.

\begin{proof}
Let us assume we have a path of states constructed by applying corresponding rules from $\mathcal{R}$ where $\mathtt{X}$ is reachable. At some point on the path, we inevitably have to create a complex agent $\mathtt{X}_2 \lhd \mathtt{X}$ from a complex agent $\mathtt{X}_1$ applying a rule $\mathtt{R}$. 

\begin{center}
\includegraphics[scale=0.15]{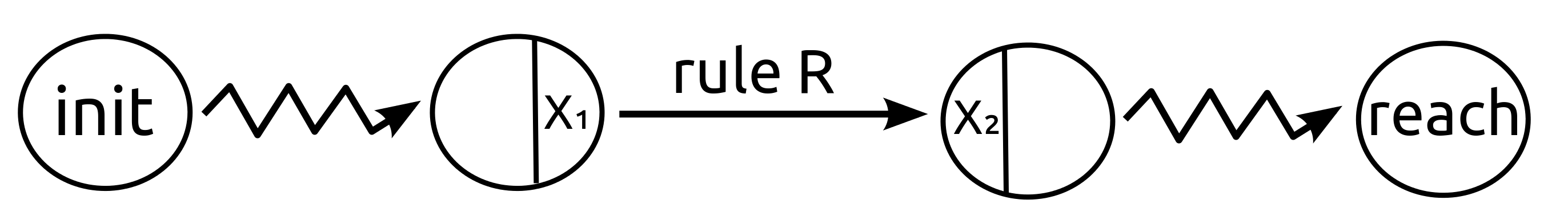}
\end{center}

It requires there has to be a complex agent $\mathtt{X}_2'$ in the rule which is compatible with the complex agent $\mathtt{X}_2$. If there is no such agent, the agent $\mathtt{X}$ is non-reachable.
\end{proof}
\end{theorem}

Compared to dynamic non-reachability analysis, Theorem~\ref{static_reach} completely avoids any combinatorial explosion and gives an answer only by checking structural properties of rules.  The usage of the theorem is demonstrated in Section~\ref{case_study}.

\section{Case study}
\label{case_study}

\begin{figure}[!h]
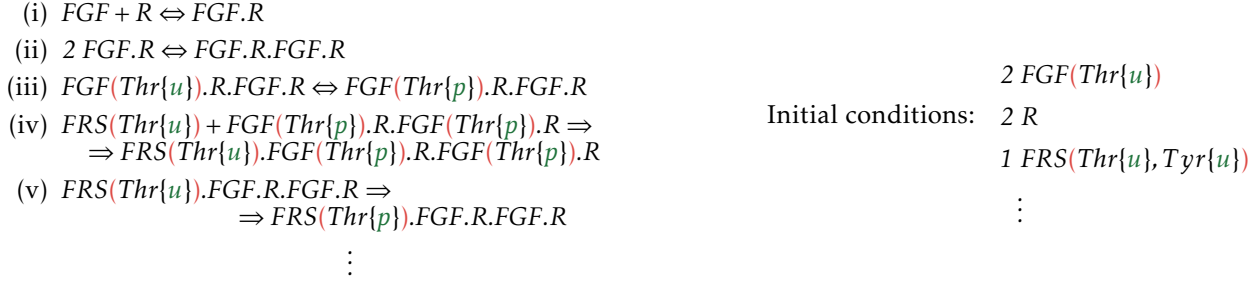

\begin{center}
\begin{minipage}{0.5\textwidth}
    \begin{enumerate}
      \item $\mathit{FGF} + R \Leftrightarrow \mathit{FGF}.R$

      \item $\mathit{2}~ \mathit{FGF}.R \Leftrightarrow \mathit{FGF}.R.\mathit{FGF}.R$

      \item $\mathit{FGF}\textcolor{myred}{(}\mathit{Thr}\{\textcolor{mydarkgreen}{u}\}\textcolor{myred}{)}.R.\mathit{FGF}.R \Leftrightarrow \mathit{FGF}\textcolor{myred}{(}\mathit{Thr}\{\textcolor{mydarkgreen}{p}\}\textcolor{myred}{)}.R.\mathit{FGF}.R$ 

      \item \label{focus_rule} $\mathit{FRS}\textcolor{myred}{(}\mathit{Thr}\{\textcolor{mydarkgreen}{u}\}\textcolor{myred}{)} + \mathit{FGF}\textcolor{myred}{(}\mathit{Thr}\{\textcolor{mydarkgreen}{p}\}\textcolor{myred}{)}.R.\mathit{FGF}\textcolor{myred}{(}\mathit{Thr}\{\textcolor{mydarkgreen}{p}\}\textcolor{myred}{)}.R \Rightarrow$ 
      
      $\Rightarrow \mathit{FRS}\textcolor{myred}{(}\mathit{Thr}\{\textcolor{mydarkgreen}{u}\}\textcolor{myred}{)}.\mathit{FGF}\textcolor{myred}{(}\mathit{Thr}\{\textcolor{mydarkgreen}{p}\}\textcolor{myred}{)}.R.\mathit{FGF}\textcolor{myred}{(}\mathit{Thr}\{\textcolor{mydarkgreen}{p}\}\textcolor{myred}{)}.R$ 
      
      \item $\mathit{FRS}\textcolor{myred}{(}\mathit{Thr}\{\textcolor{mydarkgreen}{u}\}\textcolor{myred}{)}.\mathit{FGF}.R.\mathit{FGF}.R \Rightarrow $
      
      \hspace*{2cm}$\Rightarrow \mathit{FRS}\textcolor{myred}{(}\mathit{Thr}\{\textcolor{mydarkgreen}{p}\}\textcolor{myred}{)}.\mathit{FGF}.R.\mathit{FGF}.R$ 
      
      {\centering $\vdots$ \\}

    \end{enumerate}
\end{minipage}
\hfill
\begin{minipage}{0.4\textwidth}
    Initial conditions:
    \begin{tabular}{l}
      $\mathit{2}~ \mathit{FGF}\textcolor{myred}{(}\mathit{Thr}\{\textcolor{mydarkgreen}{u}\}\textcolor{myred}{)}$\\
      $\mathit{2}~ R$\\
      $\mathit{1}~ \mathit{FRS}\textcolor{myred}{(}\mathit{Thr}\{\textcolor{mydarkgreen}{u}\}, Tyr\{\textcolor{mydarkgreen}{u}\}\textcolor{myred}{)}$\\
    \end{tabular}
    
    {\centering $\vdots$ \\}
\end{minipage}
\end{center}
  \caption{A fragment of Yamada et al. model \cite{yamada2004model} of $\mathit{FGF}$ signalling pathway written in BCSL. All agents are residing in a cytosol $cyt$ compartment, which are omitted for simplicity. The rule~(\ref{focus_rule}) requires both threonine residues ($\mathit{Thr}$) on $\mathit{FGF}$ proteins to be phosphorylated ($\textcolor{mydarkgreen}{p}$). Basically, it is not possible to create a complex from $\mathit{FRS}$ and unphosphorylated ($\textcolor{mydarkgreen}{u}$) $\mathit{FGF}$ proteins. Full model is available in Appendix~\ref{appendix}.}\label{fgf_fragment}
\end{figure}

We want to demonstrate practical purposes of static analysis defined in this paper. Yamada et al. model \cite{yamada2004model} is a model of \emph{fibroblast growth factor} (FGF) signalling pathway. The model represents a signalling pathway, which is typically a cascade of signal transduction. It means that incorrect behaviour on a particular point in the cascade will influence the rest of the pathway. The entire model written in BCSL syntax consists of 20 types of agents interacting in 57 rules. Most of proteins can undergo phosphorylation (state change from $\textcolor{mydarkgreen}{u}$ to $\textcolor{mydarkgreen}{p}$ on some amino acid residues). We consider initial conditions such that there are all required agents in one or two repetitions (in cases when there are required multiple agents to create complexes, e.g. $\mathit{FGF}$). In such case, the number of reachable states can grow up to $2^{72}$, which is too high to be effectively enumerated. In \autoref{fgf_fragment}, there is a fragment of the model required for our purposes, the whole model is available in Appendix~\ref{appendix}.

For example, we want to check whether given agent $\mathit{FRS}\textcolor{myred}{(}\mathit{Thr}\{\textcolor{mydarkgreen}{u}\}, \mathit{Tyr}\{\textcolor{mydarkgreen}{u}\}\textcolor{myred}{)}.\mathit{FGF}\textcolor{myred}{(}\mathit{Thr}\{\textcolor{mydarkgreen}{u}\}\textcolor{myred}{)}.R.\mathit{FGF}\textcolor{myred}{(}\mathit{Thr}\{\textcolor{mydarkgreen}{u}\}\textcolor{myred}{)}.R{::}cyt$ is reachable for the given model. The agent is formed from $\mathit{FGF}$ proteins which are unphosphorylated ($\textcolor{mydarkgreen}{u}$) on threonine residues ($\mathit{Thr}$). With the traditional approach, we have to enumerate entire transition system of the model and then use model checking method to check it. In our case, we can check if it is non-reachable using \emph{static reachability analysis} (Theorem~\ref{static_reach}). The conclusion is that there is no compatible agent on any right-hand side of the rules. It follows that the given complex agent is non-reachable. 

\begin{figure}[!h]
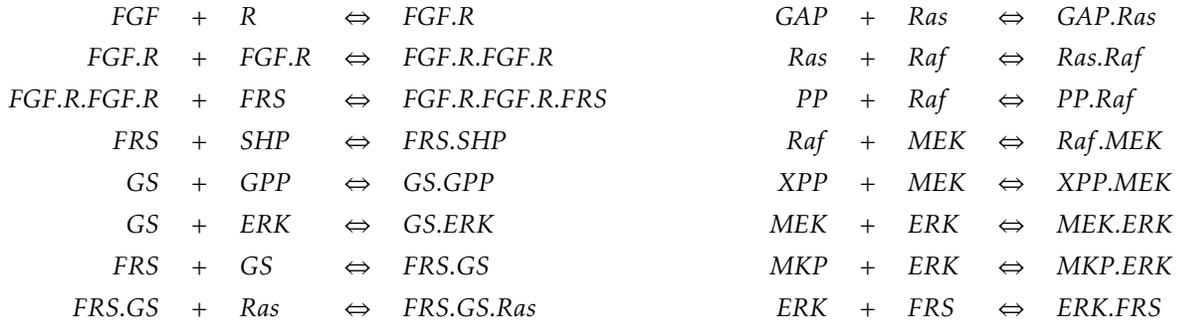

\begin{minipage}{0.45\textwidth}
      \begin{tabular}{r c l c l}
        $\mathit{FGF}$ & + & $\mathit{R}$ & $ \Leftrightarrow $ & $ \mathit{FGF.R}$\\
        $\mathit{FGF.R}$ & + & $\mathit{FGF.R}$ & $ \Leftrightarrow $ & $\mathit{FGF.R.FGF.R}$\\
        $\mathit{FGF.R.FGF.R}$ & + & $\mathit{FRS}$ & $ \Leftrightarrow $ & $\mathit{FGF.R.FGF.R.FRS}$\\
        $\mathit{FRS}$ & + & $\mathit{SHP}$ & $ \Leftrightarrow $ & $\mathit{FRS.SHP}$\\
        $\mathit{GS}$ & + & $\mathit{GPP}$ & $ \Leftrightarrow $ & $\mathit{GS.GPP}$\\
        $\mathit{GS}$ & + & $\mathit{ERK}$ & $ \Leftrightarrow $ & $\mathit{GS.ERK}$\\
        $\mathit{FRS}$ & + & $\mathit{GS}$ & $ \Leftrightarrow $ & $\mathit{FRS.GS}$\\
        $\mathit{FRS.GS}$ & + & $\mathit{Ras}$ & $ \Leftrightarrow $ & $ \mathit{FRS.GS.Ras}$\\
      \end{tabular}
\end{minipage}
\hfill
\begin{minipage}{0.4\textwidth}
      \begin{tabular}{r c l c l}
        $\mathit{GAP}$ & + & $\mathit{Ras}$ & $ \Leftrightarrow $ & $ \mathit{GAP.Ras}$\\
        $\mathit{Ras}$ & + & $\mathit{Raf}$ & $ \Leftrightarrow $ & $ \mathit{Ras.Raf}$\\
        $\mathit{PP}$ & + & $\mathit{Raf}$ & $ \Leftrightarrow $ & $ \mathit{PP.Raf}$\\
        $\mathit{Raf}$ & + & $\mathit{MEK}$ & $ \Leftrightarrow $ & $ \mathit{Raf.MEK}$\\
        $\mathit{XPP}$ & + & $\mathit{MEK}$ & $ \Leftrightarrow $ & $ \mathit{XPP.MEK}$\\
        $\mathit{MEK}$ & + & $\mathit{ERK}$ & $ \Leftrightarrow $ & $ \mathit{MEK.ERK}$\\
        $\mathit{MKP}$ & + & $\mathit{ERK}$ & $ \Leftrightarrow $ & $ \mathit{MKP.ERK}$\\
        $\mathit{ERK}$ & + & $\mathit{FRS}$ & $ \Leftrightarrow $ & $ \mathit{ERK.FRS}$\\
      \end{tabular}
\end{minipage}
  \caption{Yamada et al. model \cite{yamada2004model} after context-based reduction was applied. All agents are residing in a cytosol $cyt$ compartment, which are omitted for simplicity. Original model is available in Appendix~\ref{appendix}.}\label{reduced_yamada}
\end{figure}

Demonstration of \emph{context-based reduction} (Theorem~\ref{non-reachability_in_reduced}) is provided on the same model as in the previous case. We can compute with the entire model since we will reduce its context to the minimum. Applying the reduction, there are created 16 bidirectional rules (\autoref{reduced_yamada}). The size of transition system has significantly decreased -- it has approximately six hundreds of states and two thousands of edges.

For instance, we want check reachability of a complex agent $\mathit{Raf}\textcolor{myred}{(}Thr\{\textcolor{mydarkgreen}{p}\}\textcolor{myred}{)}.\mathit{ERK}\textcolor{myred}{(}Tyr\{\textcolor{mydarkgreen}{p}\},Thr\{\textcolor{mydarkgreen}{p}\}\textcolor{myred}{)}{::}cyt$ in the initial model. We can first check whether its corresponding least specified agent $\mathit{Raf}.\mathit{ERK}{::}cyt$ is non-reachable in the reduced model. Since the transition system of the model is relatively small, it can be quite easily checked using dynamical model checking. The answer in this case is non-reachable, which means the original agent in non-reachable too.

\section{Conclusions}
\label{conclusion}

We have presented the second generation of Biochemical Space Language, a novel high-level language for the hierarchical description of biological structures and mechanistic description of chemical reactions by means of compact rules. With respect to the previous prototype~\cite{Ded201627} the language fully utilises the specific view on the biochemical structures and reactions and the level of abstraction is not lost by translating the language into a low-level formalism not capable of maintaining a hierarchy of object types at the adequate level of abstraction.

We have defined and consequently demonstrated on several case studies static analysis techniques that are unique for the level of abstraction the language uses. We have shown it is possible to detect redundant rules and answer some reachability queries statically. The potential of the language provides the basis for further static analysis that is enabled by the specific abstraction and rule-based approach. Compared to low-level languages, we can take advantage of the hierarchy and relationships built among agents, as demonstrated in provided analysis techniques.

We are aware of necessity to deeply compare these defined relations with the concepts of other formalisms. Our notion of compatible sets has a relation to orthogonal fragments in Kappa~\cite{FERET201827}. Despite the fact that on our level of abstraction we do not have binding sites, the compatible sets can be seen as a simplified version of orthogonal fragments operating only on the level of states. Similarly, the reduction of models (and consequently reachability analysis) can be related to decontextualisation in Kappa~\cite{FeretICCMSE2007}. The formulation of exact relationships is left for the future work.

We are planning to extend the language by quantitative aspects such that we enable simulations of the models. However, this is quite a challenging task since writing a rate of the rule requires to express how particular agents from the rule participate in the rate while keeping the syntax readable and concise. We are also developing the tool BCSgen that is able to maintain and analyse BCSL specifications with its online version eBCSgen.

\bibliographystyle{entcs}
\bibliography{papers}

\appendix

\section{Model Yamada et al. 2004}
\label{appendix}

\begin{center}
\bgroup
\def\arraystretch{1.5}%
\begin{tabular}{ r c l }
$\mathit{FGF}::cyt + R::cyt $ & $\Leftrightarrow$ & $ \mathit{FGF}.R::cyt$\\
$2~ \mathit{FGF}.R::cyt $ & $\Leftrightarrow$ & $ \mathit{FGF}.R.\mathit{FGF}.R::cyt$\\
$\mathit{FGF}\textcolor{myred}{(}Thr\{\textcolor{mydarkgreen}{u}\}\textcolor{myred}{)}.R.\mathit{FGF}.R::cyt $ & $\Leftrightarrow$ & $ \mathit{FGF}\textcolor{myred}{(}Thr\{\textcolor{mydarkgreen}{p}\}\textcolor{myred}{)}.R.\mathit{FGF}.R::cyt$\\
$\mathit{FRS}\textcolor{myred}{(}Thr\{\textcolor{mydarkgreen}{u}\}\textcolor{myred}{)}.\mathit{FGF}.R.\mathit{FGF}.R::cyt $ & $\Rightarrow$ & $ \mathit{FRS}\textcolor{myred}{(}Thr\{\textcolor{mydarkgreen}{p}\}\textcolor{myred}{)}.\mathit{FGF}.R.\mathit{FGF}.R::cyt$\\
$\mathit{FRS}\textcolor{myred}{(}Thr\{\textcolor{mydarkgreen}{p}\}\textcolor{myred}{)}.\mathit{FGF}.R.\mathit{FGF}.R::cyt $ & $\Rightarrow$ & $ \mathit{FRS}\textcolor{myred}{(}Thr\{\textcolor{mydarkgreen}{p}\}\textcolor{myred}{)}::cyt + \mathit{FGF}.R.\mathit{FGF}.R::cyt$\\
$\mathit{SHP}::cyt + \mathit{FRS}\textcolor{myred}{(}Thr\{\textcolor{mydarkgreen}{p}\}\textcolor{myred}{)}::cyt$ & $\Rightarrow$ & $ \mathit{SHP}.\mathit{FRS}\textcolor{myred}{(}Thr\{\textcolor{mydarkgreen}{p}\}\textcolor{myred}{)}::cyt$\\
$\mathit{FRS}\textcolor{myred}{(}Thr\{\textcolor{mydarkgreen}{p}\}\textcolor{myred}{)}.\mathit{SHP}::cyt $ & $\Rightarrow$ & $ \mathit{FRS}\textcolor{myred}{(}Thr\{\textcolor{mydarkgreen}{u}\}\textcolor{myred}{)}.\mathit{SHP}::cyt$\\
$\mathit{FRS}\textcolor{myred}{(}Thr\{\textcolor{mydarkgreen}{u}\}\textcolor{myred}{)}.\mathit{SHP}::cyt $ & $\Rightarrow$ & $ \mathit{FRS}\textcolor{myred}{(}Thr\{\textcolor{mydarkgreen}{u}\}\textcolor{myred}{)}::cyt + \mathit{SHP}::cyt$\\
$ \mathit{GPP}::cyt + \mathit{GS}\textcolor{myred}{(}Thr\{\textcolor{mydarkgreen}{p}\}\textcolor{myred}{)}::cyt$ & $\Rightarrow$ & $ \mathit{GPP}.\mathit{GS}\textcolor{myred}{(}Thr\{\textcolor{mydarkgreen}{p}\}\textcolor{myred}{)}::cyt$\\
$\mathit{GS}\textcolor{myred}{(}Thr\{\textcolor{mydarkgreen}{p}\}\textcolor{myred}{)}.\mathit{GPP}::cyt $ & $\Rightarrow$ & $ \mathit{GS}\textcolor{myred}{(}Thr\{\textcolor{mydarkgreen}{u}\}\textcolor{myred}{)}.\mathit{GPP}::cyt$\\
$\mathit{GS}\textcolor{myred}{(}Thr\{\textcolor{mydarkgreen}{u}\}\textcolor{myred}{)}.\mathit{GPP}::cyt $ & $\Rightarrow$ & $ \mathit{GS}\textcolor{myred}{(}Thr\{\textcolor{mydarkgreen}{u}\}\textcolor{myred}{)}::cyt + \mathit{GPP}::cyt$\\
$\mathit{ERK}\textcolor{myred}{(}Tyr\{\textcolor{mydarkgreen}{p}\},Thr\{\textcolor{mydarkgreen}{p}\}\textcolor{myred}{)}::cyt + \mathit{GS}\textcolor{myred}{(}Thr\{\textcolor{mydarkgreen}{u}\}\textcolor{myred}{)}::cyt $ & $\Rightarrow$ & $ \mathit{ERK}\textcolor{myred}{(}Tyr\{\textcolor{mydarkgreen}{p}\},Thr\{\textcolor{mydarkgreen}{p}\}\textcolor{myred}{)}.\mathit{GS}\textcolor{myred}{(}Thr\{\textcolor{mydarkgreen}{u}\}\textcolor{myred}{)}::cyt$\\
$\mathit{GS}\textcolor{myred}{(}Thr\{\textcolor{mydarkgreen}{u}\}\textcolor{myred}{)}.\mathit{ERK}::cyt $ & $\Rightarrow$ & $ \mathit{GS}\textcolor{myred}{(}Thr\{\textcolor{mydarkgreen}{p}\}\textcolor{myred}{)}.\mathit{ERK}::cyt$\\
$\mathit{GS}\textcolor{myred}{(}Thr\{\textcolor{mydarkgreen}{p}\}\textcolor{myred}{)}.\mathit{ERK}::cyt $ & $\Rightarrow$ & $ \mathit{GS}\textcolor{myred}{(}Thr\{\textcolor{mydarkgreen}{p}\}\textcolor{myred}{)}::cyt + \mathit{ERK}::cyt$\\
$\mathit{FRS}\textcolor{myred}{(}Thr\{\textcolor{mydarkgreen}{p}\},Tyr\{\textcolor{mydarkgreen}{u}\}\textcolor{myred}{)}::cyt + \mathit{GS}\textcolor{myred}{(}Thr\{\textcolor{mydarkgreen}{u}\}\textcolor{myred}{)}::cyt $ & $\Leftrightarrow$ & $ \mathit{FRS}\textcolor{myred}{(}Thr\{\textcolor{mydarkgreen}{p}\},Tyr\{\textcolor{mydarkgreen}{u}\}\textcolor{myred}{)}.\mathit{GS}\textcolor{myred}{(}Thr\{\textcolor{mydarkgreen}{u}\}\textcolor{myred}{)}::cyt$\\
$\mathit{Ras}\textcolor{myred}{(}Thr\{\textcolor{mydarkgreen}{u}\}\textcolor{myred}{)}.\mathit{FRS}.\mathit{GS}::cyt $ & $\Rightarrow$ & $ \mathit{Ras}\textcolor{myred}{(}Thr\{\textcolor{mydarkgreen}{p}\}\textcolor{myred}{)}.\mathit{FRS}.\mathit{GS}::cyt$\\
$\mathit{Ras}\textcolor{myred}{(}Thr\{\textcolor{mydarkgreen}{p}\}\textcolor{myred}{)}.\mathit{FRS}.\mathit{GS}::cyt $ & $\Rightarrow$ & $ \mathit{Ras}\textcolor{myred}{(}Thr\{\textcolor{mydarkgreen}{p}\}\textcolor{myred}{)}::cyt + \mathit{FRS}.\mathit{GS}::cyt$\\
$\mathit{GAP}::cyt + \mathit{Ras}\textcolor{myred}{(}Thr\{\textcolor{mydarkgreen}{p}\}\textcolor{myred}{)}::cyt$ & $\Rightarrow$ & $ \mathit{GAP}.\mathit{Ras}\textcolor{myred}{(}Thr\{\textcolor{mydarkgreen}{p}\}\textcolor{myred}{)}::cyt$\\
$\mathit{Ras}\textcolor{myred}{(}Thr\{\textcolor{mydarkgreen}{p}\}\textcolor{myred}{)}.\mathit{GAP}::cyt $ & $\Rightarrow$ & $ \mathit{Ras}\textcolor{myred}{(}Thr\{\textcolor{mydarkgreen}{u}\}\textcolor{myred}{)}.\mathit{GAP}::cyt$\\
$\mathit{Ras}\textcolor{myred}{(}Thr\{\textcolor{mydarkgreen}{u}\}\textcolor{myred}{)}.\mathit{GAP}::cyt $ & $\Rightarrow$ & $ \mathit{Ras}\textcolor{myred}{(}Thr\{\textcolor{mydarkgreen}{u}\}\textcolor{myred}{)}::cyt + \mathit{GAP}::cyt$\\
$\mathit{Ras}\textcolor{myred}{(}Thr\{\textcolor{mydarkgreen}{p}\}\textcolor{myred}{)}::cyt + \mathit{Raf}\textcolor{myred}{(}Thr\{\textcolor{mydarkgreen}{u}\}\textcolor{myred}{)}::cyt $ & $\Rightarrow$ & $ \mathit{Ras}\textcolor{myred}{(}Thr\{\textcolor{mydarkgreen}{p}\}\textcolor{myred}{)}.\mathit{Raf}\textcolor{myred}{(}Thr\{\textcolor{mydarkgreen}{u}\}\textcolor{myred}{)}::cyt$\\
$\mathit{Raf}\textcolor{myred}{(}Thr\{\textcolor{mydarkgreen}{u}\}\textcolor{myred}{)}.\mathit{Ras}::cyt $ & $\Rightarrow$ & $ \mathit{Raf}\textcolor{myred}{(}Thr\{\textcolor{mydarkgreen}{p}\}\textcolor{myred}{)}.\mathit{Ras}::cyt$\\
$\mathit{Raf}\textcolor{myred}{(}Thr\{\textcolor{mydarkgreen}{p}\}\textcolor{myred}{)}.\mathit{Ras}::cyt $ & $\Rightarrow$ & $ \mathit{Raf}\textcolor{myred}{(}Thr\{\textcolor{mydarkgreen}{p}\}\textcolor{myred}{)}::cyt + \mathit{Ras}::cyt$\\
$\mathit{PP}::cyt + \mathit{Raf}\textcolor{myred}{(}Thr\{\textcolor{mydarkgreen}{p}\}\textcolor{myred}{)}::cyt$ & $\Rightarrow$ & $ \mathit{PP}.\mathit{Raf}\textcolor{myred}{(}Thr\{\textcolor{mydarkgreen}{p}\}\textcolor{myred}{)}::cyt$\\
$\mathit{Raf}\textcolor{myred}{(}Thr\{\textcolor{mydarkgreen}{p}\}\textcolor{myred}{)}.\mathit{PP}::cyt $ & $\Rightarrow$ & $ \mathit{Raf}\textcolor{myred}{(}Thr\{\textcolor{mydarkgreen}{u}\}\textcolor{myred}{)}.\mathit{PP}::cyt$\\
$\mathit{Raf}\textcolor{myred}{(}Thr\{\textcolor{mydarkgreen}{u}\}\textcolor{myred}{)}.\mathit{PP}::cyt $ & $\Rightarrow$ & $ \mathit{Raf}\textcolor{myred}{(}Thr\{\textcolor{mydarkgreen}{u}\}\textcolor{myred}{)}::cyt + \mathit{PP}::cyt$\\
$\mathit{Raf}\textcolor{myred}{(}Thr\{\textcolor{mydarkgreen}{p}\}\textcolor{myred}{)}::cyt + \mathit{MEK}\textcolor{myred}{(}Ser212\{\textcolor{mydarkgreen}{u}\}\textcolor{myred}{)}::cyt$ & $\Rightarrow$ & $ \mathit{Raf}\textcolor{myred}{(}Thr\{\textcolor{mydarkgreen}{p}\}\textcolor{myred}{)}.\mathit{MEK}\textcolor{myred}{(}Ser212\{\textcolor{mydarkgreen}{u}\}\textcolor{myred}{)}::cyt$\\
$\mathit{MEK}\textcolor{myred}{(}Ser212\{\textcolor{mydarkgreen}{u}\}\textcolor{myred}{)}.\mathit{Raf}::cyt $ & $\Rightarrow$ & $ \mathit{MEK}\textcolor{myred}{(}Ser212\{\textcolor{mydarkgreen}{p}\}\textcolor{myred}{)}.\mathit{Raf}::cyt$\\
$\mathit{MEK}\textcolor{myred}{(}Ser212\{\textcolor{mydarkgreen}{p}\}\textcolor{myred}{)}.\mathit{Raf}::cyt $ & $\Rightarrow$ & $ \mathit{MEK}\textcolor{myred}{(}Ser212\{\textcolor{mydarkgreen}{p}\}\textcolor{myred}{)}::cyt + \mathit{Raf}::cyt$\\
\end{tabular}
\egroup
\end{center}

\begin{center}
\bgroup
\def\arraystretch{1.5}%
\begin{tabular}{ r c l }
$\mathit{Raf}\textcolor{myred}{(}Thr\{\textcolor{mydarkgreen}{p}\}\textcolor{myred}{)}::cyt + \mathit{MEK}\textcolor{myred}{(}Ser298\{\textcolor{mydarkgreen}{u}\}\textcolor{myred}{)}::cyt$ & $\Rightarrow$ & $ \mathit{Raf}\textcolor{myred}{(}Thr\{\textcolor{mydarkgreen}{p}\}\textcolor{myred}{)}.\mathit{MEK}\textcolor{myred}{(}Ser298\{\textcolor{mydarkgreen}{u}\}\textcolor{myred}{)}::cyt$\\
$\mathit{MEK}\textcolor{myred}{(}Ser298\{\textcolor{mydarkgreen}{u}\}\textcolor{myred}{)}.\mathit{Raf}::cyt $ & $\Rightarrow$ & $ \mathit{MEK}\textcolor{myred}{(}Ser298\{\textcolor{mydarkgreen}{p}\}\textcolor{myred}{)}.\mathit{Raf}::cyt$\\
$\mathit{MEK}\textcolor{myred}{(}Ser298\{\textcolor{mydarkgreen}{p}\}\textcolor{myred}{)}.\mathit{Raf}::cyt $ & $\Rightarrow$ & $ \mathit{MEK}\textcolor{myred}{(}Ser298\{\textcolor{mydarkgreen}{p}\}\textcolor{myred}{)}::cyt + \mathit{Raf}::cyt$\\
$\mathit{XPP}::cyt + \mathit{MEK}\textcolor{myred}{(}Ser212\{\textcolor{mydarkgreen}{p}\}\textcolor{myred}{)}::cyt $ & $\Rightarrow$ & $ \mathit{XPP}.\mathit{MEK}\textcolor{myred}{(}Ser212\{\textcolor{mydarkgreen}{p}\}\textcolor{myred}{)}::cyt$\\
$\mathit{MEK}\textcolor{myred}{(}Ser212\{\textcolor{mydarkgreen}{p}\}\textcolor{myred}{)}.\mathit{XPP}::cyt $ & $\Rightarrow$ & $ \mathit{MEK}\textcolor{myred}{(}Ser212\{\textcolor{mydarkgreen}{u}\}\textcolor{myred}{)}.\mathit{XPP}::cyt$\\
$\mathit{MEK}\textcolor{myred}{(}Ser212\{\textcolor{mydarkgreen}{u}\}\textcolor{myred}{)}.\mathit{XPP}::cyt $ & $\Rightarrow$ & $ \mathit{MEK}\textcolor{myred}{(}Ser212\{\textcolor{mydarkgreen}{u}\}\textcolor{myred}{)}::cyt + \mathit{XPP}::cyt$\\
$\mathit{XPP}::cyt + \mathit{MEK}\textcolor{myred}{(}Ser298\{\textcolor{mydarkgreen}{p}\}\textcolor{myred}{)}::cyt $ & $\Rightarrow$ & $ \mathit{XPP}.\mathit{MEK}\textcolor{myred}{(}Ser298\{\textcolor{mydarkgreen}{p}\}\textcolor{myred}{)}::cyt$\\
$\mathit{MEK}\textcolor{myred}{(}Ser298\{\textcolor{mydarkgreen}{p}\}\textcolor{myred}{)}.\mathit{XPP}::cyt $ & $\Rightarrow$ & $ \mathit{MEK}\textcolor{myred}{(}Ser298\{\textcolor{mydarkgreen}{u}\}\textcolor{myred}{)}.\mathit{XPP}::cyt$\\
$\mathit{MEK}\textcolor{myred}{(}Ser298\{\textcolor{mydarkgreen}{u}\}\textcolor{myred}{)}.\mathit{XPP}::cyt $ & $\Rightarrow$ & $ \mathit{MEK}\textcolor{myred}{(}Ser298\{\textcolor{mydarkgreen}{u}\}\textcolor{myred}{)}::cyt + \mathit{XPP}::cyt$\\
$\mathit{ERK}\textcolor{myred}{(}Thr\{\textcolor{mydarkgreen}{u}\}\textcolor{myred}{)}.\mathit{MEK}::cyt $ & $\Rightarrow$ & $ \mathit{ERK}\textcolor{myred}{(}Thr\{\textcolor{mydarkgreen}{p}\}\textcolor{myred}{)}.\mathit{MEK}::cyt$\\
$\mathit{ERK}\textcolor{myred}{(}Thr\{\textcolor{mydarkgreen}{p}\}\textcolor{myred}{)}.\mathit{MEK}::cyt $ & $\Rightarrow$ & $ \mathit{ERK}\textcolor{myred}{(}Thr\{\textcolor{mydarkgreen}{p}\}\textcolor{myred}{)}::cyt + \mathit{MEK}::cyt$\\
$\mathit{ERK}\textcolor{myred}{(}Tyr\{\textcolor{mydarkgreen}{u}\}\textcolor{myred}{)}.\mathit{MEK}::cyt $ & $\Rightarrow$ & $ \mathit{ERK}\textcolor{myred}{(}Tyr\{\textcolor{mydarkgreen}{p}\}\textcolor{myred}{)}.\mathit{MEK}::cyt$\\
$\mathit{ERK}\textcolor{myred}{(}Tyr\{\textcolor{mydarkgreen}{p}\}\textcolor{myred}{)}.\mathit{MEK}::cyt $ & $\Rightarrow$ & $ \mathit{ERK}\textcolor{myred}{(}Tyr\{\textcolor{mydarkgreen}{p}\}\textcolor{myred}{)}::cyt + \mathit{MEK}::cyt$\\
$\mathit{MKP}::cyt + \mathit{ERK}\textcolor{myred}{(}Thr\{\textcolor{mydarkgreen}{p}\}\textcolor{myred}{)}::cyt $ & $\Rightarrow$ & $ \mathit{MKP}.\mathit{ERK}\textcolor{myred}{(}Thr\{\textcolor{mydarkgreen}{p}\}\textcolor{myred}{)}::cyt$\\
$\mathit{ERK}\textcolor{myred}{(}Thr\{\textcolor{mydarkgreen}{p}\}\textcolor{myred}{)}.\mathit{MKP}::cyt $ & $\Rightarrow$ & $ \mathit{ERK}\textcolor{myred}{(}Thr\{\textcolor{mydarkgreen}{u}\}\textcolor{myred}{)}.\mathit{MKP}::cyt$\\
$\mathit{ERK}\textcolor{myred}{(}Thr\{\textcolor{mydarkgreen}{u}\}\textcolor{myred}{)}.\mathit{MKP}::cyt $ & $\Rightarrow$ & $ \mathit{ERK}\textcolor{myred}{(}Thr\{\textcolor{mydarkgreen}{u}\}\textcolor{myred}{)}::cyt + \mathit{MKP}::cyt$\\
$ \mathit{MKP}::cyt + \mathit{ERK}\textcolor{myred}{(}Tyr\{\textcolor{mydarkgreen}{p}\}\textcolor{myred}{)}::cyt $ & $\Rightarrow$ & $ \mathit{MKP}.\mathit{ERK}\textcolor{myred}{(}Tyr\{\textcolor{mydarkgreen}{p}\}\textcolor{myred}{)}::cyt$\\
$\mathit{ERK}\textcolor{myred}{(}Tyr\{\textcolor{mydarkgreen}{p}\}\textcolor{myred}{)}.\mathit{MKP}::cyt $ & $\Rightarrow$ & $ \mathit{ERK}\textcolor{myred}{(}Tyr\{\textcolor{mydarkgreen}{u}\}\textcolor{myred}{)}.\mathit{MKP}::cyt$\\
$\mathit{ERK}\textcolor{myred}{(}Tyr\{\textcolor{mydarkgreen}{u}\}\textcolor{myred}{)}.\mathit{MKP}::cyt $ & $\Rightarrow$ & $ \mathit{ERK}\textcolor{myred}{(}Tyr\{\textcolor{mydarkgreen}{u}\}\textcolor{myred}{)}::cyt + \mathit{MKP}::cyt$\\
$\mathit{FRS}\textcolor{myred}{(}Tyr\{\textcolor{mydarkgreen}{u}\}\textcolor{myred}{)}.\mathit{ERK}::cyt $ & $\Rightarrow$ & $ \mathit{FRS}\textcolor{myred}{(}Thr\{\textcolor{mydarkgreen}{u}\},Tyr\{\textcolor{mydarkgreen}{p}\}\textcolor{myred}{)}.\mathit{ERK}::cyt$\\
$\mathit{FRS}\textcolor{myred}{(}Thr\{\textcolor{mydarkgreen}{u}\},Tyr\{\textcolor{mydarkgreen}{p}\}\textcolor{myred}{)}.\mathit{ERK}::cyt $ & $\Rightarrow$ & $ \mathit{FRS}\textcolor{myred}{(}Thr\{\textcolor{mydarkgreen}{u}\},Tyr\{\textcolor{mydarkgreen}{p}\}\textcolor{myred}{)}::cyt + \mathit{ERK}::cyt$\\
$\mathit{FRS}\textcolor{myred}{(}Tyr\{\textcolor{mydarkgreen}{p}\}\textcolor{myred}{)}::cyt $ & $\Rightarrow$ & $ \mathit{FRS}\textcolor{myred}{(}Tyr\{\textcolor{mydarkgreen}{u}\}\textcolor{myred}{)}::cyt$\\

$\mathit{ERK}\textcolor{myred}{(}Thr\{\textcolor{mydarkgreen}{u}\}\textcolor{myred}{)}::cyt$ & + & $\mathit{MEK}\textcolor{myred}{(}Ser212\{\textcolor{mydarkgreen}{p}\},Ser298\{\textcolor{mydarkgreen}{p}\}\textcolor{myred}{)}::cyt ~~\Rightarrow$\\
 & $\Rightarrow$ & $ \mathit{ERK}\textcolor{myred}{(}Thr\{\textcolor{mydarkgreen}{u}\}\textcolor{myred}{)}.\mathit{MEK}\textcolor{myred}{(}Ser212\{\textcolor{mydarkgreen}{p}\},Ser298\{\textcolor{mydarkgreen}{p}\}\textcolor{myred}{)}::cyt$\\
$\mathit{Ras}\textcolor{myred}{(}Thr\{\textcolor{mydarkgreen}{u}\}\textcolor{myred}{)}::cyt$ & + & $\mathit{FRS}\textcolor{myred}{(}Thr\{\textcolor{mydarkgreen}{p}\},Tyr\{\textcolor{mydarkgreen}{u}\}\textcolor{myred}{)}.\mathit{GS}\textcolor{myred}{(}Thr\{\textcolor{mydarkgreen}{u}\}\textcolor{myred}{)}::cyt ~~\Rightarrow$\\
 & $\Rightarrow$ & $ \mathit{Ras}\textcolor{myred}{(}Thr\{\textcolor{mydarkgreen}{u}\}\textcolor{myred}{)}.\mathit{FRS}\textcolor{myred}{(}Thr\{\textcolor{mydarkgreen}{p}\},Tyr\{\textcolor{mydarkgreen}{u}\}\textcolor{myred}{)}.\mathit{GS}\textcolor{myred}{(}Thr\{\textcolor{mydarkgreen}{u}\}\textcolor{myred}{)}::cyt$\\
$\mathit{FRS}\textcolor{myred}{(}Thr\{\textcolor{mydarkgreen}{u}\}\textcolor{myred}{)}::cyt$ & + & $\mathit{FGF}\textcolor{myred}{(}Thr\{\textcolor{mydarkgreen}{p}\}\textcolor{myred}{)}.R.\mathit{FGF}\textcolor{myred}{(}Thr\{\textcolor{mydarkgreen}{p}\}\textcolor{myred}{)}.R::cyt ~~\Rightarrow$\\
 & $\Rightarrow$ & $ \mathit{FRS}\textcolor{myred}{(}Thr\{\textcolor{mydarkgreen}{u}\}\textcolor{myred}{)}.\mathit{FGF}\textcolor{myred}{(}Thr\{\textcolor{mydarkgreen}{p}\}\textcolor{myred}{)}.R.\mathit{FGF}\textcolor{myred}{(}Thr\{\textcolor{mydarkgreen}{p}\}\textcolor{myred}{)}.R::cyt$\\
 $\mathit{ERK}\textcolor{myred}{(}Tyr\{\textcolor{mydarkgreen}{u}\}\textcolor{myred}{)}::cyt$ & + & $\mathit{MEK}\textcolor{myred}{(}Ser212\{\textcolor{mydarkgreen}{p}\},Ser298\{\textcolor{mydarkgreen}{p}\}\textcolor{myred}{)}::cyt ~~\Rightarrow$\\
 & $\Rightarrow$ & $ \mathit{ERK}\textcolor{myred}{(}Tyr\{\textcolor{mydarkgreen}{u}\}\textcolor{myred}{)}.\mathit{MEK}\textcolor{myred}{(}Ser212\{\textcolor{mydarkgreen}{p}\},Ser298\{\textcolor{mydarkgreen}{p}\}\textcolor{myred}{)}::cyt$\\
 $\mathit{FRS}\textcolor{myred}{(}Tyr\{\textcolor{mydarkgreen}{u}\}\textcolor{myred}{)}::cyt$ & + & $\mathit{ERK}\textcolor{myred}{(}Tyr\{\textcolor{mydarkgreen}{p}\},Thr\{\textcolor{mydarkgreen}{p}\}\textcolor{myred}{)}::cyt ~~\Rightarrow$\\
  & $\Rightarrow$ & $ \mathit{FRS}\textcolor{myred}{(}Tyr\{\textcolor{mydarkgreen}{u}\}\textcolor{myred}{)}.\mathit{ERK}\textcolor{myred}{(}Tyr\{\textcolor{mydarkgreen}{p}\},Thr\{\textcolor{mydarkgreen}{p}\}\textcolor{myred}{)}::cyt$\\
\end{tabular}
\egroup
\end{center}

\end{document}